 \documentclass[prd,aps,a4paper,twocolumn]{revtex4-1}

\usepackage{graphicx,psfrag}
\usepackage{mathrsfs}
\usepackage{amsmath,amsfonts,amssymb}
\usepackage{multirow}
\usepackage{comment}
\usepackage{bm}
\usepackage{graphicx}
\usepackage{psfrag}
\usepackage{epstopdf}
\usepackage{color}
\usepackage{mathrsfs}
\usepackage{comment}
\usepackage{multirow}
\usepackage{hyperref}
\usepackage{diagbox} 
\usepackage{amsmath,amssymb,amsthm,wasysym}
\usepackage{capt-of}
\usepackage[utf8]{inputenc}

\def\lm{{\ell m}}

\def\ha{{\hat{a}}}

\def\lm{{\ell m}}

\def\hr{{\hat{r}}}

\def\ii{{\rm i}}

\def\ha{{\hat{a}}}

\newcommand{\scri}{\mathcal{J}^+}
\newcommand{\be}{\begin{equation}}
\newcommand{\ee}{\end{equation}}

\newcommand{\RN}[1]{%
  \textup{\uppercase\expandafter{\romannumeral#1}}%
}

\definecolor{gray}{rgb}{0.5,0.5,0.5}
\definecolor{cyan}{rgb}{0,0.9,0.9}
\definecolor{orange}{rgb}{0.9,0.5,0}
\definecolor{magenta}{rgb}{1,0,1}
\definecolor{purple}{rgb}{0.8,0.4,0.8}
\definecolor{darkgreen}{rgb}{0,.6,0}
\definecolor{turquoise}{rgb}{0.25,0.88,0.82}

\begin{document}

\interfootnotelinepenalty=10000
\raggedbottom

\title{Asymptotic gravitational wave fluxes from a spinning particle  \\
        in circular equatorial orbits around a rotating black hole}

\author{Enno Harms${}^1$, Georgios Lukes-Gerakopoulos${}^2$, Sebastiano Bernuzzi${}^3$,  Alessandro Nagar${}^4$  }
\affiliation{${}^1$Theoretical Physics Institute, University of Jena, 07743 Jena, Germany}
\affiliation{${}^2$Institute of Theoretical Physics, Faculty of Mathematics and Physics, 
 Charles University in Prague, 18000 Prague, Czech Republic}
\affiliation{${}^3$DiFeST, University of Parma, and INFN, 43124, Parma, Italy}
\affiliation{${}^4$Institut des Hautes Etudes Scientifiques, 91440 Bures-sur-Yvette, France}

\begin{abstract}
We present a new computation of the asymptotic gravitational 
wave energy fluxes emitted by a {\it spinning} particle in circular 
equatorial orbits about a Kerr black hole.
The particle dynamics is computed in the pole-dipole approximation,
solving the Mathisson-Papapetrou equations with the Tulczyjew
spin-supplementary-condition.
The fluxes are computed, for the first time, by solving 
the 2+1 Teukolsky equation in the time-domain using 
hyperboloidal and horizon-penetrating coordinates.
Denoting by $M$ the black hole mass and by $\mu$ the particle mass,
we cover dimensionless background spins $a/M=(0,\pm0.9)$ and
dimensionless particle spins $-0.9\leq S/\mu^2 \leq +0.9$. 
Our results span orbits of Boyer-Lindquist coordinate radii
$4\leq r/M \leq 30$; notably, we investigate the strong-field 
regime, in some cases even beyond the last-stable-orbit.
We compare our numerical results for the
gravitational wave fluxes 
with the 2.5th order accurate post-Newtonian (PN) prediction
obtained analytically by Tanaka {\it {\it et al}.} [Phys.\ Rev.\ D 54, 3762 (1996)]:
we find an unambiguous trend of the PN-prediction 
towards the numerical results when $r$ is large.
At $r/M=30$ the fractional agreement between the full numerical 
flux, approximated as the sum over the modes $m=1,2,3$,
and the PN prediction is $\lesssim0.5\%$ in all cases tested. 
This is close to our fractional numerical accuracy ($\sim 0.2\%$).
For smaller radii, the agreement between the 2.5PN prediction and 
the numerical result progressively deteriorates, as expected. 
Our numerical data will be essential to develop suitably resummed 
expressions of PN-analytical fluxes in order to improve their 
accuracy in the strong-field regime.
\end{abstract}

\pacs{
  04.25.D-,     
  04.30.Db,   
  95.30.Sf     
  %
}

\maketitle

\section{Introduction}
\label{sec:intro}

The black hole (BH) binary problem is one of the
most interesting topics of numerical relativity
due to its relevance for gravitational wave (GW)
detection and its enormous numerical challenges.
In the test-particle limit we assume that
one of the two companions has such a small mass $\mu \ll M$,
where $M$ is the mass of the big companion,
that the system can be modeled as a fixed BH background
plus a perturbation. Thus, linearized
equations like the Regge-Wheeler-Zerilli (RWZ)
equations~\cite{Regge:1957td,Zerilli:1970se}
and the Teukolsky equation (TE)~\cite{Teukolsky:1972my,Teukolsky:1973ha} 
can be used. The particle limit is relevant in its own right
as the extreme-mass-ratio (EMR)
corner of parameter space~\cite{Barack:2003fp,Tiec:2014lba},
but even more important because,
(i)~the underlying physics are the same at any mass-ratio,
which allows to connect the particle results
to comparable mass ratios; in
some cases qualitatively~\cite{Dietrich:2014wja},
in other cases even quantitatively~\cite{Nagar:2013sga},
(ii)~semi-analytical models like the effective-one-body (EOB) 
model rely on EMR information, see, e.g.,~\cite{Damour:2009ic},
(iii)~the physical origin of experimental observations can
be disentangled with more intuition;
e.g., a clean definition of the trajectory or the couplings
of orbital and spin angular momenta is possible, and
(iv)~the computational costs are negligible
in contrast to full-fledged numerical relativity (NR)
simulations like, e.g.,~\cite{Dietrich:2015pxa}.
Hence, it is highly desirable to have tools like
a ``point-particle laboratory'';
in particular, one that allows to study
the effects of spin-spin-couplings
and spin-orbit couplings on 
the gravitational waveforms.

The treatment of point particles with spin has 
a long history in 
relativity~\cite{Frenkel1926, Lanczos1930, Mathisson:1937zz}.
A major step was Mathisson's model of a 
``gravitational skeleton''~\cite{Mathisson:2010}
(see also~\cite{tulczyjew1959motion, Steinhoff:2010zz}),
in which the energy momentum tensor of a spinning body
is expanded into its multipolar moments. Assuming
that the body be sufficiently small, one can neglect
higher multipole moments;
e.g., restricting
to the mass (monopole) and linear 
spin effects (dipole) one obtains the well-known 
``pole-dipole'' approximation.
Within this framework and using the conservation
equation $\nabla_\mu T^{\mu\nu}=0$,
Papapetrou derived his famous equations of motion (EoM)
for a spinning 
particle~\cite{Papapetrou:1951pa, Corinaldesi:1951pb}.
The nowadays standard form of these equations
was obtained
in a series of subsequent works by Tulczyjew~\cite{tulczyjew1959motion},
Dixon~\cite{Dixon:1964, Dixon:1970zza,Dixon:1970zz,dixon1974dynamics}
and Wald~\cite{Wald:1972sz},
and it is written
in terms of the world line variables
$\{v^\mu,p^\mu,S^{\mu\nu}\}$, i.e.
the tangent vector,
the linear four momentum and the spin
tensor respectively.
In this form the equations are often referred to as
the ``Mathisson-Papapetrou-Dixon'' equations (MPEQs),
emphasizing the important
early contribution of Mathisson~\cite{Mathisson:2010}
and the later reformulations of Dixon.
As discussed already
by Papapetrou, the EoM contain three degrees of 
freedom that have to be fixed in order to close the 
system of ordinary differential equations.
This freedom can be interpreted as the arbitrariness
in the choice of the reference point inside the
spinning body that shall be tracked by the EoM.
The reference point is not unique because
any spinning body has a lower bound for
its size: $R \gtrsim S/\mu$, 
where $R$ is its radius, $S$ its spin magnitude and
$\mu$ its mass~\cite{moller1949dynamique}.
Thus,
the MPEQs have to be understood as describing
a small but not pointlike body~\cite{Wald:1972sz}, 
and, in principle, one could choose any point
inside the body as the reference point that follows the EoM.
A unique choice in classical mechanics would be the center of mass,
which, however, in GR is observer dependent~\cite{Beigblboeck:1967wb}
(see~\cite{Steinhoff:2010zz} for a nice visualisation).
It is conventional to use the center of mass
with respect to which the ``spin'' of the body is defined
as the 
reference point tracked by the world line $X^\mu(\lambda)$, with $\lambda$ the proper time.
Thus, a condition that fixes the spin also fixes the reference point
traced by $X^\mu(\lambda)$,
and the freedom in the EoM can be removed by
imposing a spin-supplementary condition (SSC).
Several such SSCs
have been proposed in the
literature~\cite{Pryce:1948pf, Newton:1949cq, Corinaldesi:1951pb,
Pirani:1956tn, Dixon:1964,tulczyjew1959motion,Kyrian:2007zz},
usually by demanding the reference-point to coincide
with the center-of-mass as perceived by some preferred
timelike observer
and, consequently, with the spin to be orthogonal
to this observer's direction of motion.

Over the years the dynamics of a pole-dipole particle
was studied in great detail within the 
Mathisson-Papapetrou framework and under the influence
of different SSCs,
see, e.g.,~\cite{Semerak:1999qc,Kyrian:2007zz, Bini:2000vv,Hartl:2002ig,Hartl:2003da}
for an overview. Note, that 
technically the ``gravitational skeleton''
can be used to derive EoM at any
multipolar order and especially second (quadrupole) moments
modeled as quadratic in spin effects
have already been included in many 
recent works, see, e.g.,
\cite{Steinhoff:2010zz,dixon1974dynamics,Ehlers1977,
Hinderer:2013uwa, Bini:2008zzf, Bini:2013uwa, Bini:2014wua, Bini:2014epa}.
On the contrary, the literature on GWs emitted by 
spinning particles that move in BH spacetimes is 
rather sparse; probably, due to the fact that the 
interest in the topic is mostly theoretical.
Estimates on the honest effect
of the spin of the particle predict at best secular
relevance in intermediate-mass-ratios~\cite{Semerak:1999qc,Huerta:2011kt}. 
Therefore, it does not come as a surprise that all studies
mentioned below went beyond the range of astrophysically
realistic values for the spins (see Ch.~\RN{2}, Sec.~3B in~\cite{Hartl:2002ig}).
This is motivated by the hope
that high spin values can nonetheless provide 
very valuable theoretical information.

Almost 20 years ago Mino {\it et al}.~\cite{Mino:1995fm}
performed first numerical studies of the GW emission
from a spinning particle on
radial plunges along the rotation
axis of a Kerr BH
by solving the TE in the frequency domain,
more precisely within the Sasaki-Nakamura
formalism~\cite{Sasaki:1981sx}.
The work of Mino {\it et al}.~\cite{Mino:1995fm}
was complemented by radial plunges in the equatorial plane,
as considered by Saijo {\it et al}.\ in~\cite{Saijo:1998mn}.
Recently, Han~\cite{Han:2010tp}
computed numerically energy fluxes of a spinning particle in
circular, equatorial orbit about a rotating Kerr BH, both
to infinity and down the horizon, using the same approach
in the frequency-domain.
In a somewhat different approach,
using metric perturbations and linearizing the 
Einstein equations,
Tominaga {\it et al}.~\cite{Tominaga:2000cs,Tominaga:1999iy}
investigated GWs from a spinning particle
moving in the spherically symmetric background
created by a neutron star.
In 2015, Burko and Khanna~\cite{Burko:2015sqa}
studied the effect of
the particle spin on the waveforms produced
along circular motion around a Schwarzschild BH and
found that it can be as important as conservative self-force effects.
All the studies mentioned so far,
restricted the spin of the central BH, here denoted $\vec{S}_1$,
and the spin of the small body, $\vec{S}_2$,
to be (anti-)parallel
- and we will do so, as well.
On the analytical side, PN calculations
for spinning binaries of comparable masses were performed
in~\cite{Kidder:1992fr,Kidder:1995zr,Blanchet:1995ez}
and in several works since, see~\cite{Steinhoff:2010zz} for an overview.
In the point particle limit, Tanaka {\it et al}.~\cite{Tanaka:1996ht} 
computed energy fluxes to infinity at 2.5PN by solving the 
Sasaki-Nakamura equations.
Their results, Eqs~(5.17) and Eqs~(5.19) in~\cite{Tanaka:1996ht},
comprise the next-to-leading-order (NLO) in the spin-orbit interaction and 
the leading-order (LO) in the spin-square interaction
and will serve as the target solutions for our study.
Recently, new analytical expressions for the fluxes became 
available at higher PN accuracy
(in both the spin-orbit and the spin-spin interaction),
remarkably in the comparable-mass case: spin-orbit 
contributions to the flux were obtained at next-to-next-to-leading order in 
Ref.~\cite{Bohe:2013cla,Marsat:2013wwa},
see also~\cite{Blanchet:2011zv,Blanchet:2012sm}, while spin-square terms 
were given at NLO in Ref.~\cite{Bohe:2015ana},
notably also with the addition of the 
4PN tail contribution~\cite{Marsat:2013caa}.

For simplicity, here we choose to not include the test-particle limit of the 
analytical expression of Refs.~\cite{Marsat:2013wwa,Bohe:2015ana},
i.e.\ we only rely on the 2.5PN accurate original expression
of Tanaka {\it et al}.~\cite{Tanaka:1996ht}. 
The higher-order PN terms will be included, in 
a suitably factorized and resummed form~\cite{Damour:2014sva},
in a follow-up work that
aims at assessing in detail the accuracy of the analytical expressions
in the strong-field regime.

In this work we will solve the MPEQs with the Tulczyjew-SSC
(see below) for circular equatorial orbits, using 
a variational Gauss-Runge-Kutta integration scheme as presented
in~\cite{Lukes-Gerakopoulos:2014dma}.
The obtained dynamics will be fed to 
a time-domain Teukolsky solver,
which was 
successfully used in the computation of tail
decay rates~\cite{Harms:2013ib} and
GWs from EOB-radiation-reaction-driven
particle inspirals~\cite{Harms:2014dqa, Nagar:2014kha}.
The obtained GWs are used to compute energy fluxes
to infinity.
We compare our numerical data
against the 2.5PN result of
Tanaka {\it et al}.~\cite{Tanaka:1996ht}.
We will prove that our data approach towards the PN-prediction
as $r \to \infty$ until the differences reach the level
of our numerical uncertainty estimate, which thus
mutually confirms both our numerical implementation and
the analytical PN calculation of Tanaka {\it et al}.\ (also
the test-particle limit of the comparable-mass PN calculation 
of~\cite{Blanchet:2011zv,Blanchet:2012sm,Bohe:2013cla,Marsat:2013caa}). 
Moreover,
we will set up a database with numerical values for the
infinity fluxes,
which will serve as an orientation for future
studies and, in particular, allow to assess the success of
manipulations in the analytical formulas.
Note that in the
nonspinning case the PN-results have been successfully processed
in~\cite{Damour:2008gu,Pan:2010hz,Fujita:2014eta}
with resummation techniques in order to extend the
range of accuracy towards the strong-field regime.

The paper is organized as follows. In Sec.~\ref{sec:dynamics}
we will shortly describe the MPEQs and the procedure
of finding circular orbits. In Sec.~\ref{sec:teukolsky} we
review the TE and our approach to 
solving it numerically in the time-domain. In addition, we
present our strategy of computing the TE source term
with spin and the implementation in the {\texttt{Teukode}}.
Notably, our code is, to the best of
our knowledge, the first one
to solve the TE in the time-domain with a source term
for a spinning particle.
In Sec.~\ref{sec:results}
a discussion of our numerical results
for the total energy fluxes of a spinning particle
is given,
with respect to the analytical 2.5PN prediction.
Finally, in Appendix~\ref{app:CompInfos}
we complement the consistency checks of
our numerics with 
(i)~an analysis of the multipolarly decomposed fluxes,
(ii)~repeating the comparison for the total
flux with respect to a different PN variable
than that used in the main body, and
(iii)~a comparison against the numerical results
of~\cite{Han:2010tp}.
In Appendix~\ref{app:Tables}
we collect our results in tables, as a reference
for future studies.

We use geometric units, $c=G=1$, and 
the Riemann tensor defined as
${R^\alpha}_{\beta\gamma\delta}=
 \Gamma^\alpha_{\gamma \lambda} \Gamma^\lambda_{\delta \beta}
 - \partial_\delta \Gamma^\alpha_{\gamma\beta}
 - \Gamma^\alpha_{\delta\lambda} \Gamma^{\lambda}_{\gamma\beta}
 + \partial_\gamma \Gamma^{\alpha}_{\delta \beta}$,
where the Christoffel symbols $\Gamma$ are computed
from the metric with signature $(-,+,+,+)$. 
We often employ reduced
variables as denoted with a hat;
e.g., $\hr=r/M$ and $\ha=a/M=\pm|\vec{S}_1|/M^2$,
where $\vec{S}_1$ is the angular momentum,
$M$ the mass of the central BH and the
$+\, (-)$ is chosen when $\vec{S}_1$ is aligned (antialigned)
with the strictly positive orbital angular momentum.
The spin of the particle is expressed here, as well
as in most of the literature on this subject, 
with the dimensionless quantity
$\sigma=S/{(\mu M)}= \pm|\vec{S}_2|/{(\mu M)}$, where
$\vec{S}_2$ is the spin angular momentum 
of the particle and $\mu$ its conserved mass.
If we think of the particle as a model of a BH,
its maximal spin angular momentum would be $S=\pm\mu^2$
(see \cite{Hartl:2002ig} for a discussion on maximal 
spins of stellar objects), which means $\sigma$ 
would be restricted to $- \nu \leq \sigma \leq \nu$,
with $\nu\equiv \mu/M$, that is $-1\leq \sigma/\nu\leq 1$
(see also~\cite{Faye:2006gx} for a related discussion).
In practice, we use $M=\mu=1$ in 
our numerics, avoiding the inconvenient
appearance of factors $\nu$ and restricting
our spin parameter to $\sigma \in [-1,1]$.
This setting is somewhat counterintuitive
to the condition $\mu\ll M$, which we assume
in doing perturbation theory, but since $\mu$
and $M$ are just scales in the used equations
we are free to use the most convenient values numerically.

\section{Dynamics of a Pole-Dipole Particle}
\label{sec:dynamics}

In this Section we briefly recall the EoM
for a pole-dipole particle and explain
our choice for the SSC. More details on the
used numerical integration scheme can be
found in~\cite{Lukes-Gerakopoulos:2014dma}, 
where one of us compared
the dynamics of the MPEQs
with the EoM of the Hamiltonian for
a spinning particle~\cite{Barausse:2009aa}.
In addition,
we will outline the procedure
for finding initial data that leads to
circular equatorial orbits (CEOs),
including unstable ones.
We mention that, equivalently, the dynamics
for a pole-dipole particle under the TUL-SSC
which moves on a circular orbit in the equatorial
plane with (anti-)aligned spin can be computed analytically following the
Appendix of~\cite{Tanaka:1996ht}.

\subsection{Equations of Motion and the Spin-Supplementary Condition}

The MPEQs have been the object of many studies,
see, e.g.,~\cite{Semerak:1999qc} for a review
and~\cite{Hackmann:2014tga,Bini:2014soa,Semerak:2015dza}
for recent works. In the nowadays standard form
they read
\begin{align} 
 \label{eq:MPEQs}
 v^\alpha \nabla_\alpha \; p^\mu & = 
 - \dfrac{1}{2} \; {R^\mu}_{\nu\rho\sigma} \; v^\nu \; S^{\rho\sigma} \\ \nonumber
 v^\alpha \nabla_\alpha \; S^{\mu\nu} & = 
 p^\mu v^\nu - p^\nu v^\mu \qquad \qquad ,
\end{align}
where $v^\mu=d X^\mu/d\lambda$ is the tangent vector to the world line
$X^\mu(\lambda)$, with $\lambda$ the proper time, of a
yet unspecified reference point inside the spinning body and $p^\mu$
its total linear four-momentum. The spin tensor $S^{\mu\nu}$
is defined in some analogy to classical spin angular momentum
via spatial integrals over the stress-energy tensor $T^{\mu\nu}$ 
on a given hypersurface and with respect to the chosen
reference point~\cite{Kyrian:2007zz,Hartl:2002ig}. 
The more intuitive picture of 
spin as a three-dimensional vector can be partially retrieved
after the reference point is fixed through a SSC (see below).
Independently from the SSC,
any background symmetry implies a constant of motion with respect
to evolution upon the MPEQs.
Namely, for 
a Killing vector $\xi^\mu$ the quantity 
\begin{align}
 C=\xi^\mu p_\mu-\frac12 \xi_{\mu;\nu} S^{\mu\nu}
\end{align}
remains conserved upon evolution.

As mentioned earlier, the system of
Eqs.~\eqref{eq:MPEQs} contains three degrees of 
freedom associated with the arbitrariness
of the tracked reference point and thus with the notion of \textit{spin}.
This arbitrariness has to be removed 
before one can use Eqs.~\eqref{eq:MPEQs} to find
dynamics of a pole-dipole particle.
Here, we choose to close the system
with the ``Tulczyjew-SSC''~\cite{tulczyjew1959motion}
(TUL-SSC)
\begin{align}
 S^{\mu\nu} p_\mu = 0  \quad ,
\end{align}
which is known to feature conservation of the
dynamical rest mass $\mu:=\sqrt{-p^\mu p_\mu}$ and of
the spin magnitude 
\begin{align}
 \label{eq:spinMagnitude}
 S^2= \frac{1}{2} S^{\mu\nu}S_{\mu\nu} \; .
\end{align}
Note that all previous numerical studies on the topic
of energy fluxes from spinning particles made the same choice 
\cite{Mino:1995fm,Saijo:1998mn,Han:2010tp,Tanaka:1996ht}.
For the TUL-SSC a spin four-vector is defined as
\begin{align}
\label{eq:SpinVect}
 S_\mu = -\frac{1}{2} \epsilon_{\mu\nu\rho\sigma}
          \, u^\nu \, S^{\rho\sigma} \qquad,
\end{align}
where $u^\nu := p^\nu/\mu$ is the specific four momentum,
$\epsilon_{\mu\nu\rho\sigma}=\sqrt{-g} \tilde{\epsilon}_{\mu\nu\rho\sigma} $
is the Levi-Civita tensor with 
the Levi-Civita symbol $\tilde{\epsilon}_{0123}=1$ and
$g$ is the determinant of the background metric tensor.
Eq.~\eqref{eq:SpinVect} can be rearranged to get
\begin{align}
   S^{\rho\sigma}=-\epsilon^{\rho\sigma\gamma\delta} S_{\gamma} u_\delta \qquad. 
   \label{eq:T4VSin}
\end{align}
Substituting Eq.~(\ref{eq:T4VSin}) 
into the definition of the spin-magnitude, Eq.~(\ref{eq:spinMagnitude}),
we get that the spin magnitude can be written in terms of the 
four-vector
\begin{align}
\label{eq:SpinCons}
 S^2=S^\mu S_\mu \; .
\end{align}
The TUL-SSC by construction implies the orthogonality of the spin-vector
and the linear momentum four-vector, $S_\mu p^\mu=0$,
and it has been shown that $S_\mu v^\mu=0$ holds as well,
see, e.g., \cite{Hackmann:2014tga}.

In general, the imposition of a SSC closes the system of evolution
equations for the world line variables
$\{v^\mu,p^\mu,S^{\mu\nu}\}$.
For the TUL-SSC one can deduce an explicit relation
$v^\mu(u^\mu,S^{\mu\nu})$; namely,
\begin{align}
 \label{eq:v_p_TUL}
 v^\mu = \frac{\textsf{m}}{\mu} \left(
          u^\mu + 
          \frac{ 2 \; S^{\mu\nu} R_{\nu\rho\kappa\lambda} u^\rho S^{\kappa\lambda}}
          {4 \mu^2 + R_{\alpha\beta\gamma\delta} S^{\alpha\beta} S^{\gamma\delta} }
          \right)  \qquad ,
\end{align}
where $\textsf{m}=-p_\mu v^\mu$ is the rest mass with respect
to $v^\mu$, which for the TUL-SSC is only conserved
up to linear order in $S$.
In practice, the value of $\textsf{m}$ is set such that
the tangent to the world line satisfies the condition $v^\mu~v_\mu=-1$.
Interestingly, relation~\eqref{eq:v_p_TUL}
does not, in general, obey $v^\mu \parallel u^\mu$
but rather
$ v^\mu = u^\mu + \mathcal{O}(S^2)$, i.e.\ 
the specific linear momentum and the tangent vector
differ by a quadratic-in-spin term.
We mention that the relevant 
literature~\cite{Mino:1995fm,Saijo:1998mn,Han:2010tp,Tanaka:1996ht}
that we use for comparisons also uses the TUL-SSC.

Here, it is useful
to comment on the implications of the quadratic-in-spin
term introduced to the system by
virtue of the TUL-SSC.
The region of interest in this paper is the weak-field,
where $R_{\alpha\beta\gamma\delta} \approx 0$ so that
the spin-square term in Eq.~\eqref{eq:v_p_TUL} is suppressed and our dynamics
remain basically linear-in-spin (which directly reflects in the obtained GW fluxes).
Instead, in the strong-field and at large spin values $\sigma\sim\mathcal{O}(1)$
the $\mathcal{O}(S^2)$-terms have a relevant influence
on the dynamics~\footnote{See also~\cite{Lukes-Gerakopoulos:2014dma}, where one of us
  studied systematically the differences
  between the MPEQs and the strictly linearized
  Hamiltonian approach of~\cite{Barausse:2009aa}
  and found deviations at high spin values}.
Let us briefly discuss what this entails
for the significance of our study in the strong-field; especially,
with respect to the question of how reliable
our results might be for more realistic bodies with
a \textit{small} but nonvanishing quadrupolar moment.
The very first assumption of the pole-dipole model
is that the energy-momentum tensor of the
body that we aim to model is assumed
to exhibit only zeroth and first moments
when subjected to a multipolar expansion.
When this assumption is strictly satisfied,
the maintenance of the $\mathcal{O}(S^2)$-terms
introduced by the TUL-SSC is fully compatible
with the pole-dipole approximation.
On the contrary, if one was to stretch the 
limits of the model beyond a rigorous description,
by considering a realistic body for which the assumption of vanishing
higher multipole moments is violated to 
a small extent,
the pole-dipole approach becomes \textit{ab initio}
an approximation which neglects quadratic-in-spin terms
(because in realistic bodies the
spin induces a quadrupole moment which is described
as a spin-square term~\cite{Bini:2015zya,Steinhoff:2012rw,Steinhoff:2010zz}).
Thus, for such bodies the pole-dipole dynamics can only give a qualitative
description of the leading-order in $S$ behaviour.
To at least consistently investigate this leading-order in $S$
influence on the motion,
one would have to neglect quadratic-in-spin
terms all over, also in Eq.~\eqref{eq:v_p_TUL}.
In the future, we plan on doing so by repeating our analysis
with the MPEQs linearised in the spin in order
to have dynamics which are fully consistent
for realistic bodies also in the strong-field
(though restricted to linear order in the spin).

%
%
In summary, we have decided to solve
the full quadratic-in-spin relation,
Eq.~(\ref{eq:v_p_TUL}).
Thus, our dynamics is rigorous, also in the strong-field, as long as the 
considered body can in fact be described only 
by its monopole and dipole.
Instead, bodies with a small but
nonvanishing quadrupole moment are only
in the weak-field consistently described
by our dynamics - and only at leading order in $S$;
in the strong-field, the sustained $\mathcal{O}(S^2)$-term
in the dynamics is inconsistent for such realistic bodies.
Astrophysically relevant objects like BHs in a binary
are in general not consistently described by our dynamics
at small orbital distances
because we neglect the expected nonvanishing
quadrupolar moments that are induced by the spin and by tidal
deformations, see, e.g.,~\cite{Thorne:1980ru}.
Before applying our results to modeling fluxes from
realistic bodies in the strong-field, one would thus
need to repeat our analysis with consistent, linear-in-spin
dynamics to estimate the influence of the $\mathcal{O}(S^2)$-term.
Fortunately, for $\sigma\ll1$ the discussion
is anyway redundant because in that case one can at any rate
neglect all quadratic-in-spin terms;
in particular, as already argued by 
Tulczyjew \cite{tulczyjew1959motion,Dixon:1964},
one would then linearize Eq.~\eqref{eq:v_p_TUL}, which is
frequently seen in analytical approaches
like \cite{Faye:2006gx,Bini:2014ica}.

\subsection{The Kerr spacetime background}

The Kerr spacetime in Boyer-Lindquist (BL) coordinates $(t,r,\theta,\phi)$ reads
 \begin{eqnarray}
  ds^2 &=& g_{tt}~dt^2+2~g_{t\phi}~dt~d\phi + g_{\phi\phi}~d\phi^2 \nonumber \\
       &+& g_{rr}~dr^2+g_{\theta\theta}~d\theta^2 \qquad, \label{eq:LinEl}
 \end{eqnarray}
 where
 \begin{eqnarray}
   g_{tt} &=&-1+\frac{2 M r}{\Sigma}~~,\nonumber\\ 
   g_{t\phi} &=& -\frac{2 a M r \sin^2{\theta}}{\Sigma}~~,\nonumber\\
   g_{\phi\phi} &=& \frac{\Lambda \sin^2{\theta}}{\Sigma} \qquad, \label{eq:KerrMetric}\\
   g_{rr} &=& \frac{\Sigma}{\Delta}~~,\nonumber\\
   g_{\theta\theta} &=& \Sigma~~,\nonumber
 \end{eqnarray} 
 and
 \begin{eqnarray}
  \Sigma &=& r^2+ a^2 \cos^2{\theta}~~,\nonumber\\
  \Delta &=& \varpi^2-2 M r~~,\nonumber \\ 
  \varpi^2 &=& r^2+a^2~~, \nonumber \\ 
  \Lambda &=& \varpi^4-a^2\Delta \sin^2\theta \qquad.  \label{eq:Kerrfunc} 
 \end{eqnarray}

For a stationary and axisymmetric background like the Kerr one, we have two
Killing vector fields, $\xi^\mu_{(t)}=\delta^\mu_t$ and
$\xi^\mu_{(\phi)}=\delta^\mu_\phi$ respectively. The first Killing vector
provides the conserved energy
 \begin{align}\label{eq:EnCons}
  E &= -p_t+\frac12g_{t\mu,\nu}S^{\mu\nu} \qquad,
 \end{align}
while the latter provides the conserved component along the 
symmetry axis $z$ of the total angular momentum,
 \begin{align}\label{eq:JzCons}
  J_z &= p_\phi-\frac12g_{\phi\mu,\nu}S^{\mu\nu} \qquad. 
 \end{align}

\subsection{Circular Equatorial Orbits}

\label{subsec:CEOdynamics}

 The procedure to find circular equatorial orbits (CEO) for a spinning particle 
 on the Kerr spacetime background follows the guidelines given in \cite{Hackmann:2014tga}.
 Namely, it is assumed that the only nonzero component of the spin vector
 \eqref{eq:SpinVect} is the polar one, i.e.,
 \begin{align}\label{eq:SpinAli}
  S^\mu=S^\theta \delta_\theta^\mu \qquad,
 \end{align}
 This assumption together with the orthogonality conditions $S_\mu~p^\mu=0$, $S_a~v^\mu=0$ 
 leads to
 \begin{align}
  p^\theta=v^\theta=0 \qquad.
 \end{align}
 Since $v^\theta=0$, we can choose to stay on the equatorial plane, i.e., $\theta=\pi/2$.
 The condition \eqref{eq:SpinAli} on the equatorial plane means that the spin of
 the particle is parallel to the spin of the central black hole. 
 For
 $a~S>0$ the spins of the binary are aligned while for 
 $a~S<0$ the spins are antialigned. 
 
 From Eqs.~\eqref{eq:SpinCons} and \eqref{eq:SpinAli} we get $S_\theta=-\sqrt{g_{\theta\theta}}~S$.
 The only nonzero components of the spin tensor computed from Eq.~\eqref{eq:T4VSin}
 are 
 \begin{align}
  S^{tr} &= -S~u_\phi \sqrt{-\frac{g_{\theta\theta}}{g}}= -S^{rt}  \qquad, \nonumber \\
  S^{t\phi} &= S~u_r \sqrt{-\frac{g_{\theta\theta}}{g}}= -S^{\phi t}  \qquad, \nonumber \\
  S^{r\phi} &= -S~u_t \sqrt{-\frac{g_{\theta\theta}}{g}}= -S^{\phi r}  \; .
 \end{align}
 
 We can express $p_t$ and $p_\phi$ as functions of the energy $E$ and of the
 $z$-component of the total angular momentum, $J_z$,
 by using Eqs.~\eqref{eq:EnCons} and \eqref{eq:JzCons},
 and, since the system is constrained to the equatorial plane, we get 
 \begin{align}
  p_t &=\frac{-E-\dfrac{M S}{\mu r^3}(a E-J_z)}{1-\dfrac{M~S^2}{\mu^2 r^3}}\qquad, \nonumber \\
  p_\phi &=\frac{J_z-\dfrac{a M S}{\mu r^3}\left[\left(-1+\dfrac{r^3}{a^2 M}\right)a E+J_z\right]}
   {1-\dfrac{M~S^2}{\mu^2 r^3}} \; ,
 \end{align}
 cf.\ Eqs.~(40) and~(41) 
 in~\cite{Hackmann:2014tga}
 (our definitions of $E$ and $J_z$,
  Eqs.~\eqref{eq:EnCons} and \eqref{eq:JzCons},
  differ from those of~\cite{Hackmann:2014tga}, Eqs.~(37) and (38),
  by a minus sign).
 Rewriting the above expressions in dimensionless quantities,
 we have
 \begin{align}
  \label{eq:specific_linmom_CEO_TUL}
  u_t &=\frac{-\hat{E}-\dfrac{\sigma}{\hat{r}^3}(\hat{a}\hat{E}-\hat{J}_z)}{1-\dfrac{\sigma^2}{\hat{r}^3}}\qquad, \nonumber \\
  u_\phi &= M \frac{\hat{J}_z-\dfrac{\hat{a} \sigma}{\hat{r}^3}\left[\left(-1+\dfrac{\hat{r}^3}{\hat{a}^2}\right)
  \hat{a} \hat{E}+\hat{J}_z\right]}{1-\dfrac{\sigma^2}{\hat{r}^3}} \qquad,
 \end{align}  
 where  $ \hat{J}_z:=J_z/(\mu M)$, and $ \hat{E}:=E/\mu$. 
 
 From Eq.~\eqref{eq:v_p_TUL} and Eq.~\eqref{eq:specific_linmom_CEO_TUL}
 we can get the radial velocity as
 a function of the dimensionless quantities as well, i.e., 
 \begin{align} \label{eq:RadVel}
   \frac{d\hat{r}}{d \hat{\lambda}}=\frac{(\sigma^2-\hat{r}^3)^3\sqrt{V_\textrm{eff}}}{\hat{r}~ Q} \qquad,
 \end{align}
 where $\hat{\lambda}=\lambda/M$, 
 \begin{align*}   
Q & = \hat{r}^{12}-4 \hat{r}^9 \sigma ^2-6 \hat{r}^7 \sigma ^2 (\hat{J_z}-\hat{E} (\hat{a}+\sigma ))^2 \\
  & +6 \hat{r}^6 \sigma ^4
-3 \hat{r}^4 \sigma ^4 (\hat{J_z}-\hat{E} (\hat{a}+\sigma ))^2-4 \hat{r}^3 \sigma ^6+\sigma ^8 \; ,
 \end{align*}
  and the effective potential
 \begin{widetext}
 \begin{align}
V_{\rm eff}&=\left(\hat{E}^2-1\right) \hr^8 +  2 \hr^7
    +\hr^6 \left[\ha^2 \left(\hat{E}^2-1\right)-(\hat{E} \sigma-\hat{J}_z)^2\right]+2 \hr^5 \left[\ha\hat{E} (\ha \hat{E}+3 \hat{E} \sigma-2 \hat{J}_z)+(\hat{E} \sigma-\hat{J}_z)^2+\sigma (\sigma-\hat{E} \hat{J}_z)\right]\nonumber\\
       & - 4 \hr^4 \sigma^2 + 2 \ha \sigma  \hr^3\left[(\ha^2+\ha \sigma) \hat{E}^2 -\hat{E}\hat{J}_z(\sigma+2\hat{a}) +\ha \sigma +\hat{J}_z^2\right]
    +\hr^2 \sigma^2 \left[(\ha \hat{E}-\hat{J}_z)^2-\sigma^2\right]
   +2 \hr \sigma^4 -\ha^2 \sigma^4~~ \, .
 \end{align}   
 \end{widetext}
 
 The turning points of the radial motion, defined by $V_\textrm{eff}=0$,
 provide the first condition to obtain CEOs. The second condition
 is to demand that
 there is no radial acceleration.
 Altogether,
 \begin{align} \label{eq:VeffCEO}
  V_\textrm{eff} = 0, \qquad \frac{d V_\textrm{eff}}{d \hat{r}} = 0 \; ,
 \end{align}
 have to be fulfilled for a CEO.
 By solving the system of Eqs.~\eqref{eq:VeffCEO} for a given radial distance $\hat{r}$
 and spin $\sigma$ we get the energy $\hat{E}$, and the component of the total
 angular momentum $\hat{J}_z$. If $\frac{d^2 V_\textrm{eff}}{d \hat{r}^2} < 0$,
 the CEO is stable along the radial direction while, if 
 $\frac{d^2 V_\textrm{eff}}{d \hat{r}^2} > 0$, the CEO is unstable. In the latter
 case, we have to impose in the code that the radial and the polar velocity are
 zero during the whole evolution of the system, otherwise small numerical
 instabilities will drive the numerical trajectory off the CEO.
For all values of the (background and particle) spins considered in this work, 
we computed the radius of the last stable orbit (LSO)
by demanding $\frac{d^2}{d\hr^2} V_{\rm{eff}} =0$ in addition to
Eqs.~\eqref{eq:VeffCEO}
(see e.g. \cite{Suzuki:1997by,Hackmann:2014tga} for the same
calculation). The corresponding numbers are listed in Table~\ref{tab:LSOs},
together with the respective orbital frequency of the particle
as computed from Eq.~\eqref{eq:Kepler}.

\begin{table}[t]
\caption{ 
Characterization of the last stable orbit (LSO) of a spinning particle
on circular equatorial orbits of a Kerr background under 
the MPEQs and the TUL-SSC, Eqs.~\eqref{eq:MPEQs} and~\eqref{eq:v_p_TUL}.
The columns report, for each value of $(\hat{a},\sigma)$ considered,
the dimensionless radius of the LSO, $\hat{r}_{\rm LSO}$,
and the corresponding dimensionless orbital frequency, $\hat{\Omega}_{\rm LSO}$,
obtained from Eq.~\eqref{eq:Kepler}.}
\centering
\begin{ruledtabular}
  \begin{tabular}[t]{c | c c c| c c c }
 &  
 \multicolumn{3}{c|}{ {\bf\large{$\hat{r}_{\rm{LSO}}$}} }  &
 \multicolumn{3}{c}{ {\bf\large{$\hat{\Omega}_{\rm{LSO}}$}} } \\ 
   \hline
   \backslashbox{$\sigma$}{$\ha$}
     &  $-0.9$  &    0.0     &  $+0.9$ &  $-0.9$  &    0.0     &  $+0.9$ \\ 
  \hline
  +0.9 &  6.606 & 4.083 & 1.663  & 5.572e-02 & 1.049e-01 & 3.123e-01\\
  +0.7 &  7.209 & 4.603 & 1.755  & 5.030e-02 & 9.178e-02 & 2.938e-01\\
  +0.5 &  7.710 & 5.063 & 1.873  & 4.649e-02 & 8.243e-02 & 2.748e-01\\
  +0.3 &  8.147 & 5.470 & 2.025  & 4.357e-02 & 7.553e-02 & 2.552e-01\\
  +0.1 &  8.536 & 5.833 & 2.216  & 4.126e-02 & 7.024e-02 & 2.350e-01\\
  0.0 &  8.717 & 6.000 & 2.321  & 4.026e-02 & 6.804e-02 & 2.254e-01\\
  $-0.1$ &  8.890 & 6.160 & 2.429  &3.935e-02 & 6.606e-02 & 2.165e-01\\
  $-0.3$ &  9.216 & 6.457 & 2.644  &3.774e-02 & 6.267e-02 & 2.009e-01\\
  $-0.5$ &  9.517 & 6.729 & 2.841  &3.637e-02 & 5.988e-02 & 1.891e-01\\
  $-0.7$ &  9.799 & 6.981 & 3.011  &3.517e-02 & 5.751e-02 & 1.808e-01\\
  $-0.9$ &  10.062& 7.214 & 3.147  &3.412e-02 & 5.551e-02 & 1.758e-01\\
  \end{tabular} 
 \end{ruledtabular}
\label{tab:LSOs}
\end{table}

 \subsection{Orbital configurations}

  We consider orbits with Boyer-Lindquist radii 
  $\hat{r}=\{5,6,7,8,10,12,15,20\}$. For some 
  configurations, further data points at $\hat{r}=4$ 
  and $\hat{r}=30$ are computed. Comparing
  with Table~\ref{tab:LSOs}, this means we
  are considering also unstable orbits beyond the LSO.
  For each value of $\hat{r}$ we consider 
  three values of the background spin
  $\hat{a}=\{-0.9,0,+0.9\}$ and ten values of the particle spin
  $\sigma=\{-0.9,-0.7,-0.5.-0.3,-0.1,+0.1,+0.3,+0.5,+0.7,+0.9\}$.
  
  After obtaining the corresponding initial
  data following Sec.~\ref{subsec:CEOdynamics},
  we numerically integrate the EoM,
  Eq.~\eqref{eq:MPEQs} and Eq.~\eqref{eq:v_p_TUL}.
  As an immediate check of our dynamics
  we compare the obtained numerical frequency
  $\Omega^{\rm Num}= v^\phi/v^t$ with the exact
  analytical expression,
  Eq.~(12) in~\cite{Han:2010tp},
  that generalizes the usual Kepler law from geodesic motion
  to our EoM for a spinning particle
  (cf.\ also the Appendix of~\cite{Tanaka:1996ht}, which gives
  a procedure to compute the frequency as well).
  Within our notation, this (dimensionless) orbital frequency reads
\begin{widetext}
  \begin{align}
\label{eq:Kepler}
   \hat{\Omega}\equiv M \Omega = \frac{
   \ha +
   \sigma  \left(\frac{3}{2} + 3 \ha^2 u^2\right)+\frac{1}{2} \sigma^2 \ha (3 + 4 u) u^2
   -u^{-3/2}
   \sqrt{
    1 +
    3 \ha \sigma  u^2 +
    \frac{13}{4} \sigma ^2 u^3 +
    \frac{3}{2} \ha \sigma ^3 u^5 +
   \left(\frac{9}{4} \ha^2 u^7-2 u^6\right) \sigma ^4}
   }   
   { \ha^2 - u^{-3} + 3 \sigma  \left(\ha^3 u^2+\ha\right)+\sigma ^2 \left[1 + \ha^2 (2 u+3) u^2\right]  },
  \end{align}
\end{widetext}
where $u\equiv 1/\hat{r}=M/r$. We use here the minus sign in the $\mp$ 
case distinction in Eq.~(12) of~\cite{Han:2011qz} because we define 
$-1\leq \ha\leq 1$ (thus a case distinction is not needed). 
In the $\sigma\to 0$ limit, this equation reduces to the usual Kepler
constraint for a nonspinning particle on Kerr background,
$\hat{\Omega}=1/(\hat{a}+\hat{r}^{3/2})$.
Also, when working at linear order in $\sigma$, it coincides with
the PN expression given by Tanaka {\it et al}.~\cite{Tanaka:1996ht}, see 
Eq.~\eqref{eq:PN_Omega} below.

The dimensionless orbital frequency $\hat{\Omega}^{\rm Num}$ obtained from 
direct numerical integration of the EoM, Eqs.~\eqref{eq:MPEQs} and~\eqref{eq:v_p_TUL},
is found to agree with the analytical expression Eq.~\eqref{eq:Kepler}
within a fractional difference of $10^{-4}$  or smaller. 
We take this as a reliable cross check of the numerical 
implementation of the dynamics. As a consequence, in the following we will just use
Eq.~\eqref{eq:Kepler} above whenever showing results
as a function of the PN-ordering parameter $x\equiv \hat{\Omega}^{2/3}$.

\section{Teukolsky Formalism in the Time-Domain}
\label{sec:teukolsky}

In this Section we review our approach
to solve the TE in the time domain by means 
of the \texttt{Teukode}
(see~\cite{Harms:2013ib,Harms:2014dqa,Nagar:2014kha} for further details).
We discuss a strategy of 
computing the source term of the TE
for a pole-dipole particle.
We spare a detailed repetition of the description
of the \texttt{Teukode},
only mentioning that it was validated
for a nonspinning particle
by reproducing several previous
results~\cite{Bernuzzi:2012ku,Bernuzzi:2011aj,Bernuzzi:2010xj,
Barausse:2011kb,Sundararajan:2010sr,Sundararajan:2007jg,
Hughes:1999bq,Taracchini:2014zpa}.

\subsection{(2+1) Approach and HH-coordinates}

Our approach to solving the gravitational TE
makes use of hyperboloidal, horizon penetrating
``HH-coordinates'' $\{\tau,\rho,\theta,\varphi\}$,
see~\cite{Calabrese:2005rs,Zenginoglu:2007jw,Zenginoglu:2009hd,Zenginoglu:2010cq,Bernuzzi:2011aj,Vano-Vinuales:2014koa}
for general ideas and~\cite{Yang:2013uba,Harms:2014dqa} for explicit formulas.
The HH-coordinates smoothly reach future null infinity $\mathcal{J}^+$ (``scri''), at $\rho_S$,
and penetrate the horizon at $\rho_+$, which makes them favorable for numerical computations
in two aspects: 
(i)~the domain $[\rho_+,\rho_S]$
is causally closed with vanishing radial coordinate light speeds
at the boundaries (thus no numerical boundary conditions have to be imposed), and
(ii)~$\rho_+$ and $\scri$ constitute the
two most interesting points for GW-extraction.
Since these points are part of the computational domain,
we avoid the additional uncertainties of
extrapolation procedures.
As pointed out in~\cite{Harms:2013ib,Harms:2014dqa},
the application of these coordinates
to the TE has to go hand in hand with a change of the underlying tetrad
for the resulting equation to be regular at the horizon and at
$\scri$. A rotation of the tetrad is
equivalent to a rescaling of the field $\Psi \rightarrow \psi$,
see~\cite{Harms:2014dqa} for the explicit rescaling implied
by switching from the Kinnersley-tetrad to the
Campanelli-tetrad~\cite{Campanelli:2000nc},
which is used here.
The azimuthal coordinate $\varphi$ is the one from
the ingoing-Kerr coordinate system and thus adopted
to the axisymmetry of the Kerr spacetime,
i.e. $\partial_\varphi = \partial_\phi$,
where $\phi$ is the
standard BL-coordinate.
Consequently, a decomposition into Fourier $m$-modes,
$\psi = \sum_m \psi_m e^{\ii m \varphi}$,
results in components $\psi_m$ of the
full solution that decouple upon evolution
and can be treated separately.
In practice, rederiving the TE in the tetrad of~\cite{Campanelli:2000nc}
with components specified in the HH-coordinate
system and separating the azimuthal dependence,
we obtain a reformulated (2+1)-TE of the form 
\begin{align}
 \label{eq:TE} 
 &  C_{\tau\tau} \partial_{\tau\tau}\psi_m 
 +  C_{\tau\rho} \partial_{\tau\rho} \psi_m 
 +  C_{\rho\rho}  \partial_{\rho\rho}\psi_m 
 +  C_{\theta\theta}  \partial_{\theta\theta}\psi_m \nonumber \\
 + &C_{\tau}  \partial_\tau \psi_m
 +  C_{\theta} \partial_{\theta}\psi_m 
 +  C_{\rho}  \partial_\rho\psi_m  
 +  C_{0} \psi_m = S_s \ , 
\end{align}
with coefficients $C(\rho,\theta; m,s)$ depending on the background
coordinates, the spin weight $s$, and the azimuthal mode-index $m$. 
Below, the index $m$ in the complex variable $\psi_m$
will be suppressed for brevity.

\subsection{The Source Term}

The TE, Eq.~\eqref{eq:TE}, contains
the source term $S_s$,
which encodes the specific form 
of the matter perturbation that
shall be treated.
For the gravitational cases the source terms
are given by
\begin{align}
\label{eq:S-2}
S_{-2} &= 8 \pi \Sigma (r- \ii a\cos\theta)^4 T_4 \ , \\
\label{eq:S+2}
S_{+2} &= 8 \pi \Sigma T_0 \ .
\end{align}
The tetrad scalars
$T_4$, $T_0$ are built from the stress-energy tensor of
the perturbation $T^{\mu\nu}$ and from
the legs $l^\mu,n^\mu,m^\mu$
of the tetrad, which was used for the derivation
of Eq.~\eqref{eq:TE}.
Schematically,
$T_4$, and similarly $T_0$, is computed as
\begin{align}
 \label{eq:T4definition}
 T_{4} = \mathcal{D}_1 \left( T^{\mu\nu} n_\mu n_\nu \right)
        + \mathcal{D}_2 \left( T^{\mu\nu} m_\mu m_\nu \right)
        + \mathcal{D}_3 \left( T^{\mu\nu} n_\mu m_\nu \right)
  ,
\end{align}
where the $\mathcal{D}_i$ are certain sums of all kinds of
1st and 2nd coordinate-derivative operators
and of non-derivative algebraic factors
(see~\cite{Teukolsky:1973ha} for details). 

To be consistent with our dynamics, the model for the 
stress-energy tensor of a spinning particle must
be Mathisson's ``gravitational skeleton'' in the
pole-dipole approximation
\begin{align}  
\label{eq:Tmunu_dipole}
 \sqrt{-g} T^{\mu\nu}  &=  \int d\lambda \; 
  \left\{
   t^{\mu\nu}(\lambda)  \delta^4\left(x^\mu-X^\mu(\lambda)\right)  \nonumber  \right. \\
   & \left.  - \nabla_\alpha \left[
                  t^{\mu\nu\alpha}(\lambda) \delta^4\left(x^\mu-X^\mu(\lambda)\right)
                    \right]   
  \right\}    \qquad ,
\end{align}
where $x^\mu=\{\tau,x^i\}$ are some coordinates
(in our case the HH-coordinates), $\lambda$ is the proper time,
$X^\mu(\lambda)$ is the world line of the reference point inside 
the modeled body, $t^{\mu\nu}=v^{(\mu}p^{\nu)}$ is the monopole moment
and $t^{\mu\nu\alpha}=S^{\alpha(\mu}v^{\nu)}$ is the dipole moment.
Transforming the integral via 
$ d\lambda \rightarrow dX^0/ \left( \frac{d}{d\lambda} X^0(\lambda) \right)$
and exploiting the defining properties of a $\delta$-distribution,
we obtain the ready-to-use expression 
(cf. Eqs.~(2.14) and (2.16) of~\cite{Faye:2006gx})
\begin{align}
  \label{eq:Tmunu_poledipole}
   T^{\mu\nu}(\tau,x^i)  = \frac{1}{\sqrt{-g}}
    \left\{ 
     \frac{v^{(\mu}p^{\nu)}}{v^t} \delta^3 
      - \nabla_\alpha \left( \frac{ S^{\alpha(\mu}v^{\nu)}} {v^t} \delta^3 \right)    
    \right\} \, ,
\end{align}
where $\delta^3$ abbreviates $\delta^3\left(x^i-X^i(\tau)\right)$.
In Eq.~\eqref{eq:Tmunu_poledipole}
the dynamical quantities, which
are obtained from solving the MPEQs,
are depending on the background-coordinate-time,
i.e.\ 
$X^i(\tau),v^\mu(\tau),p^\mu(\tau),S^{\mu\nu}(\tau)$,
but not on the spatial coordinates.
The factor $\sqrt{-g}$ 
and the Christoffel symbols,
which enter through the covariant derivative in Eq.~\eqref{eq:Tmunu_dipole},
are functions of the background and not coordinate time-dependent,
at least in time-symmetry adapted coordinates like the HH-
or the BL-coordinates.
Since the TE source 
term is computed through Eq.~\eqref{eq:T4definition}
from all kinds of 
$(\tau,\rho,\theta,\varphi)$-derivatives of $T^{\mu\nu}$,
a consistent handling of these dependencies
is important.

Unfortunately, the explained ``choices'' for the dependencies
are not unique. The appearance of 
$\delta$-distributions without any integrals
in Eq.~(\ref{eq:Tmunu_poledipole}) creates room for ambiguity.
If Eq.~(\ref{eq:Tmunu_poledipole}) was under an integral $\int dx^i (\dots)$,
there would be no doubt that we were free to interchange
field and source points, $x^i \leftrightarrow X^i(\tau)$,
by virtue of the $\delta$-distributions.
But the integral is missing so that we have to consider the
$\delta$-distributions as usual functions that only approximate
the distributions in some limit.
Therefore, we decide to avoid,
(i)~interchanging of $x^i$ and $X^i$ at will, and
(ii)~shifting derivatives that hit the $\delta$-function
to the remaining integrand by virtue of 
$\int dx f(x) \partial_x \delta(x-y) = - \int dx \ \delta(x-y) \partial_x f(x)$.
While these choices are certainly
the most reasonable in our opinion, one could still
argue that the $\delta$-functions act as if there was an integral.
In particular, we mention that
reference \cite{Faye:2006gx} explicitly uses the Christoffel symbols
at $X^i(t)$ (working with BL-time $t$),
thus making them $t$-dependent instead of $x^i$-dependent.
In principle, this leads to slight differences compared to our choice when
evaluating the TE-source term.
However, a good approximation of the
$\delta$-distribution means that $\delta(x^i)\neq0$ only
if $x^i\approx X^i(t)$,
which, after all, means the whole discussion
becomes irrelevant
at high enough resolutions, at least with respect to the numerical outcome.

Let us proceed by discussing briefly the actual implementation
of the source term.
Since the different derivative terms in
Eq.~(\ref{eq:T4definition}) turn out to be
of large algebraic complexity,
it is advisable to wrap single smaller parts into 
auxiliary quantities. For instance, one starts by
separating the part that in the case of $S^{\mu\nu}=0$
reduces to the nonspinning particle
energy-momentum tensor multiplied by $\sqrt{-g}$,
$\tilde{T}^{\mu\nu}_{\rm{NS}}= v^{(\mu}p^{\nu)} \delta^3 / v^t$,
from the explicitly $S^{\mu\nu}$-dependent part
$\tilde{T}^{\mu\nu}_{\rm{SP}}$, which reads
\begin{align}
 \sqrt{-g} \, T^{\mu\nu} = \tilde{T}^{\mu\nu}_{\rm{NS}} + \tilde{T}^{\mu\nu}_{\rm{SP}}  \; .
\end{align}
Note though that the rewriting
$\tilde{T}^{\mu\nu}_{\rm{NS}}= \mu v^{\mu} v^{\nu} \delta^3 /v^t$,
which we used in~\cite{Harms:2014dqa},
does not hold anymore because
we have $p^\mu \neq \mu v^\mu$ if $S^{\mu\nu}\neq0$.
We further separate the explicitly spin-dependent part into
pieces, according to
\begin{align}
\tilde{T}^{\mu\nu}_{\rm{SP}} &= Q_A^{\mu\nu} + Q_B^{\mu\nu} + Q_C^{\mu\nu} +Q_D^{\mu\nu} \qquad , \\
 Q_A^{\mu\nu} & :=  \partial_\tau 
          \left( S^{0(\mu} V^{\nu)} \; \delta^3  \right)   \;\; , \nonumber \\   
 Q_B^{\mu\nu} & :=  S^{i(\mu} V^{\nu)} \; \partial_i \delta^3 \;\; , \nonumber \\
 Q_C^{\mu\nu} & :=  S^{\rho\mu} \, V^\sigma \Gamma^\nu_{\rho\sigma} \; \delta^3 \;\; ,\nonumber \\
 Q_D^{\mu\nu} & :=  S^{\rho\nu} \, V^\sigma \Gamma^\mu_{\rho\sigma} \; \delta^3 \;\; , \nonumber
\end{align}
where we have introduced the coordinate velocities
$V^\mu = v^\mu/v^\tau$ and exploited the antisymmetry of $S^{\mu\nu}$.
In the code, the derivatives of $\tilde{T}^{\mu\nu}_{\rm{SP}}$ are then
computed as sums of derivatives of the $Q_{X}^{\mu\nu}$.
Notably, $Q_A^{\mu\nu}$ and
$Q_B^{\mu\nu}$ already contain $\partial\delta$-terms,
which means that we will end up with a source term
that includes third derivatives of the $\delta$-functions,
i.e.\ $\partial\partial\partial \delta$-terms.
It remains to compute derivatives
of the tetrad legs and, finally, to shuffle all the
pieces together.

At last, we emphasize that in the \texttt{Teukode}
the whole computation outlined above is performed
in the HH-coordinate system. This is not necessarily
to be expected since $T_0$ and $T_4$ are tetrad scalars
and thus coordinate invariants, but it turns out
that the $\delta$-function treatment is simplified
if the source is treated within the same coordinates
as the rest of the equation, see Sec.~3.2 in \cite{Harms:2014dqa}.
This means, after solving
the MPEQs in Boyer-Lindquist coordinates,
we have to transform $\{X^\mu,v^\mu,p^\mu,S^{\mu\nu}\}$
before feeding them to the code.

\subsection{Numerical Approximation of $\delta$-functions}

\label{subsec:Delta_funs}

The performance of different numerical approximations of a 
$\delta$-distribution, and of its derivatives in $r$ and $\theta$, 
was discussed already in~\cite{Harms:2014dqa}. 
However, when the particle is spinning
the presence of third derivatives of $\delta$-functions
in the source term obliges us
to reconsider what is the best option.
Inspecting the recent literature, one finds that a $\delta$-like
source can be implemented essentially in two ways: 
(i)~a few-point (discrete) approximation,
developed in~\cite{Tornberg:2004JCoPh.200..462T,Engquist:2005JCoPh.207...28E,Sundararajan:2007jg},
and 
(ii)~a simpler, analytical approximation by a (narrow) Gaussian
function.
Within our numerical infrastructure, the two representations for
the $\delta$-function were extensively compared in~\cite{Harms:2014dqa}.
It was found that the few-point representation is superior
to the extended representation for CEOs. 
In that case we were able to reproduce the values
for energy fluxes emitted by a nonspinning particle as 
computed by S.~Hughes by means of an improved version of
the frequency-domain code used 
in~\cite{Hughes:1999bq,Hughes:2001jr,Sundararajan:2007jg}
(see Appendix A of \cite{Taracchini:2013wfa})
within $0.01\%$ at a resolution of 
$N_\rho \times N_\theta = 2400 \times 200$ points.
In general, however, for noncircular motion
the Gaussian approximation of the $\delta$-function
turned out to be the better option as it caused
less numerical noise in the simulation.

In the present study of CEOs we use 
the narrow-Gaussian model because,
i)~it will be our standard choice when deviating from
circular motion in the future, and
ii)~the prescriptions for building a few-point representation
of third derivatives of a delta-function have not been derived in~\cite{Sundararajan:2007jg}
(though following the same ideas they can be in principle).
This approximation implies
a slight loss of accuracy compared to our results 
for a nonspinning particle~\cite{Harms:2014dqa}.
The fractional accuracy 
we obtain is about $0.2\%$ in the dominant 
$m=2$ mode (see below), which is sufficient to prove 
the consistency between the PN and the numerical fluxes 
at large orbital radii.

\section{Results: comparing numerical and post-Newtonian fluxes for circular orbits}
\label{sec:results}
In this Section we present our new numerical
computation of the GW energy fluxes
emitted by a spinning particle on a
CEO about a Kerr black hole.
We discuss the total flux to infinity~\footnote{We have
  also obtained preliminary results for horizon fluxes, 
  which will be discussed in a separate study.},
which is approximated here as the sum over
the three dominant $m=1,2,3$ multipoles,
with all corresponding $\ell$-contributions included,
and compare it with the 2.5PN prediction.
We see that the result of the numerical computation 
and the analytical PN prediction are consistent (up to the $0.2\%$ level) 
for large orbital radii, i.e. when $x\equiv \hat{\Omega}^{2/3} \to 0$.
A detailed multipolar analysis is presented in Appendix~\ref{app:CompInfos}.

We mention the possibly obvious fact that our
results hold analogously
for angular momentum fluxes.
For circular orbits the energy fluxes and the angular momentum fluxes
are trivially connected via 
$ \frac{dJ}{dt} = \frac{m}{\omega} \; \frac{dE}{dt}$,
(see Eq.~(4.13) in~\cite{Teukolsky:1974yv}).
Here, the orbital frequency of the particle, $\Omega$,
determines the waveform's frequency $\omega= m \Omega$.

\subsection{Numerical fluxes and their accuracy}
\label{sub:nums}

In this section we will give an estimate of the
accuracy of our energy flux computations 
on the basis of \textit{non}-spinning particle
experiments, which can be 
quantitatively compared with existing
literature results.
More precisely, we compute the fluxes of
a \textit{non}-spinning particle
on a CEO,
with the same numerical setup that 
will be used for a spinning particle,
i.e.\ using the Gaussian approximation of the $\delta$,
and then we compare with the extremely 
accurate target solution computed by S.~Hughes
in the frequency domain~\cite{Hughes:1999bq,Hughes:2001jr,Sundararajan:2007jg,Taracchini:2013wfa}.

In~\cite{Harms:2014dqa},
we have performed the same comparison 
and found a $\sim0.01 \%$ agreement
at resolutions $N_\rho \times N_\theta =2400 \times 200$.
However, this remarkable accuracy relies
on the usage of
the few-point delta-approximation of~\cite{Sundararajan:2007jg}.
The Gaussian $\delta$-approximation, which we use here,
is expected to entail a loss of accuracy.
To compensate the loss of accuracy,
we increase the resolution in this
study to $N_\rho \times N_\theta =4800 \times 400$ points.
As shown below, this is enough to reach satisfactory
accuracy for $\hr\leq20$.
When studying orbits at $\hr=30$,
the accuracy drops due to the 
loss of resolution near the compactification boundary
in the hyperboloidal coordinate system.
Therefore, at $\hr=30$ we employ extraordinary
resolutions of $N_\rho \times N_\theta =6000 \times 500$.

Before going into detail on the accuracy,
we discuss how the energy fluxes 
can be computed from our master variable $\psi$.
At future null infinity, $\rho=\rho_S$,
our master variable satisfies
$ \psi(\tau,\rho_S,\theta,\varphi) = r \Psi_4(\tau,\theta,\varphi)$.
Therefore, we can compute 
the multipolar energy fluxes to infinity
directly from $\psi$, which for the multipolar decomposition reads
\begin{align}
 \label{eq:Eflux}
  F_{\lm} = \frac{2}{16 \pi} | r \dot{h}_\lm |^2
          = \frac{2}{4 \pi \omega^2}  | \psi_\lm |^2 \; ,
\end{align}
where $r \dot{h} = 2 \int \psi dt'$.
The multipolar energy flux $F_\lm$ is here defined to include 
both the $+m$ and $-m$ contributions, which
implies the factor $2$ in Eq.~\eqref{eq:Eflux},
cf.~Eq.~(2) of~\cite{Damour:2008gu}.
Note that the $\lm$-subscript refers to a multipolar
decomposition that can be done either with 
respect to the spin-weighted spherical harmonics $Y_{s\lm}$
or the spin-weighted spheroidal harmonics $S_{s\lm}$,
see App.~\ref{sec:Sslm} for a short description of these functions.
Notably, the reference data for $\sigma=0.0$ includes
the multipolar fluxes with respect to both 
the $Y_{2\lm}$ and the $S_{-2\lm}$ bases.

Let us proceed by
comparing our numerical results 
and the target solution
of Hughes for $\sigma=0.0$.
As a general bench mark,
we find an agreement of $\sim 0.2\%$
in the full $\ell$-summed flux of a $m=2$ simulation
for all tested background spins $\ha \in \{ 0,\pm0.9 \}$,
and at all radii, $\hr=(4,5,6,7,8,10,12,15,20,30)$.
We find approximately the same accuracy
in the dominant multipolar fluxes.
For example, at $\hr=20$ for $\ha=\pm0.9$
the multipolar $22$-flux, either
with respect to $Y_{-222}$ or to $S_{-222}$,
exhibits again a $\sim0.2\%$ agreement with the
reference solution.
For the subdominant modes
the accuracy decreases slightly
because they are much smaller
in absolute magnitude;
e.g., for both the $Y_{-232}$-flux and the
$S_{-232}$-flux we find a $\sim0.5\%$
agreement in the $\hr=20$, $\ha=\pm0.9$ test cases.
In conclusion,
the estimated $\lesssim 0.2-0.5\%$-level of accuracy
for $\sigma=0.0$ 
is surely enough to prove consistency
with the corresponding PN expressions,
though nonzero values of the spin,
$\sigma\neq0$, are expected to slightly increase
the amount of noise in our simulations.
To get an immediate impression of the
accuracy of our multipolar fluxes, we have added the nonspinning 
$\sigma=0.0$ results of \cite{Hughes:1999bq}
to our plots of the multipolar
fluxes over $\sigma$, Fig.~\ref{fig:Eflux_scr_Slm_NNxr_overS}.
The $\sigma=0.0$ data points are highlighted by short 
horizontal gray lines, which should be cut by the
smooth connection of our $\sigma \neq0$ data points.

It is now necessary
to point out a further practical detail
regarding our flux computations.
The two relations in Eq.~\eqref{eq:Eflux},
i.e.\ the computation in terms of $\dot{h}$
and in terms of $\psi$ respectively,
are analytically equivalent, but in practice
we find a crucial numerical difference at large radii.
At small radii, we find that both computations are
numerically indifferent, as expected.
For example, for $\ha=-0.9$ at $\hr=5$ 
we find an agreement between both
computations of $\sim 0.00001\%$, independently from
the considered value of $\sigma$.
At larger radii the agreement between the two
ways to compute the fluxes gets progressively worse;
for example, at $\hr=20$ it amounts to $\sim 0.01\%$,
which is still equivalent for our purpose 
from a practical point of view.
But, at $\hr=30$ it can make a significant difference
whether the fluxes are computed from $\psi$ directly
or from $\dot{h}$ after integration.
For example, doing a test at $\hr=30$ with $\ha=0.9$ and $\sigma=0.0$,
the computation using $\dot{h}$
works reliably - 
we find a deviation in the full $m=2$ flux
(including all $\ell$ contributions)
from the target solution of only $\sim 0.1 \%$.
On the contrary,
the computation of fluxes 
directly from $\psi$ is more biased by numerical noise
- we find, e.g., for the flux in the $Y_{21}$-mode
a deviation of $\sim1.8\%$ from the target.
For $\sigma\neq0$ the level of noise can be even worse so that
we cannot in all cases extract an unambiguous energy flux
directly from $\psi$.

The better accuracy in case of using $\dot{h}$
for the flux computation originates from
the cumulative integration:
it acts like a filter for the numerical noise,
which is present in all our simulations
since we do not employ any numerical dissipation.
At small radii, $\hr \leq 20$ 
the absolute values of our fields
are so large that the noise is at a negligible level.
At large radii, $\hr > 20$, however, the amplitude 
of the fields becomes so small 
that the numerical noise can spoil 
the accuracy of our computations.
In summary, we are able to extract
energy fluxes at large radii, $\hr>20$, in a robust way
when using a numerical integration of $\psi$ to obtain $\dot{h}$.
Unfortunately, in the data sets that we had compiled for this study
these integrations were only performed for the
$Y_{-2\lm}$-fluxes, whereas the
$S_{-2\lm}$-fluxes were computed using
$\psi$ directly and are thus only stated
up to $\hr=20$. In the future, another possibility
is to use numerical dissipation, which is
implemented though not used for this study
because our initial tests at small radii
did not suggest that it might be needed.

Finally, as another interesting side remark,
we mention that
the fluxes in the $(2,2),(2,1),(3,3)$-modes
are practically equal when doing the
multipolar decomposition either with respect 
to $Y_{-2\lm}$ or to $S_{-2\lm}$.
For example, for $\ha=\pm0.9$ at $\hr=20$
the $Y_{-222}$-flux and the $S_{-222}$-flux
differ by only $\sim0.01 \%$.
Similarly, at this configuration
the $Y_{-221}$-flux and the $S_{-221}$-flux,
and the $Y_{-233}$-flux and the $S_{-233}$,
differ by $\sim0.02\%$.
This agreement is sustained, to large extent,
even at strong-field radii, where the
frequencies - and thus the differences
between the $Y$-functions and the $S$-functions -
are larger.
For example, at $\hr=5$
the $Y_{-222}$-flux and the $S_{-222}$
flux differ by $\sim0.2\%$ for $\ha=+0.9$,
and by $\sim0.5\%$ for $\ha=-0.9$.
In summary, for these modes
any discussion held in terms of a
$S_{-2\lm}$-decomposition holds
to large extent as well for a $Y_{-2\lm}$-decomposition.
On the contrary,
the fluxes in the $(3,2)$-mode are drastically
different in the two basis-systems.
For example, at $\hr=20$ for $\sigma=0.0$
and $\ha=0.9$ we find deviations of
$\sim 60\%$ (!).

\subsection{Analytics: Post-Newtonian energy fluxes}
\label{sub:pn}
Tanaka {\it et al}.~\cite{Tanaka:1996ht} computed, long ago, in post-Newtonian
theory, the GW energy fluxes emitted by a spinning 
particle on a CEO about a Kerr black hole.
Their results hold at $2.5$PN-order,
and are linear in the particle spin,
i.e.\ including only leading-order (LO) spin contributions.
The respective formulas are expressed using the dimensionless spin magnitude
(called $\hat{s}$ in~\cite{Tanaka:1996ht}).
The main result of~\cite{Tanaka:1996ht}
is the total energy flux to infinity,
which reads, in terms of $u\equiv M/r$, where
$r$ is the BL-coordinate radius, 
\begin{align}
\label{eq:F_bl}
F =\dfrac{32}{5}\nu^2 u^5&\bigg\{1 - \dfrac{1247}{336}u + \left(4\pi - \dfrac{73}{12} \ha -\dfrac{25}{4}\sigma\right)u^{3/2}\nonumber\\
  &+\left(-\dfrac{44711}{9072}+\dfrac{33}{16}\ha^2 + \dfrac{71}{8}\ha\sigma\right)u^2\nonumber\\ 
 &+\left(-\dfrac{8191}{672}\pi + \dfrac{3749}{336}\ha + \dfrac{2403}{112}\sigma\right)u^{5/2}\bigg\},
\end{align}
where $\nu\equiv \mu/M$ is the mass ratio.
Note that in~\cite{Tanaka:1996ht} the notation $v^2=M/r$ was used, 
which we will \textit{not} follow here in view of
the fact that $M/r$ is not the orbital velocity (except for $\ha=\sigma=0$).
To drive comparisons with numerical data,
as well as with other PN results,
one has to recast the above expression
into a formulation in 
terms of the gauge-invariant standard
PN-ordering parameter $x\equiv \hat{\Omega}^{2/3}$.
Note that, for nonzero values
of $(\hat{a},\sigma)$, one has $x\neq u$,
i.e\ $\hat{\Omega}\neq u^{3/2}$. 
To rewrite Eq.~\eqref{eq:F_bl}, first in terms of $\hat{\Omega}$
and subsequently in terms of $x$,
one can PN-expand the analytic expression
for the frequency, Eq.~\eqref{eq:Kepler}, 
which yields
\be
\label{eq:PN_Omega}
\hat{\Omega} = u^{3/2}\left[1-\left(\dfrac{3}{2}\sigma+\ha\right)u^{3/2}+\dfrac{3}{2}\ha\sigma u^2 +O(u^3)\right],
\ee
see Eq.~(5.18) in~\cite{Tanaka:1996ht}.
We invert this equation,
insert it into Eq.~\eqref{eq:F_bl}, and, adopting the notation
\begin{align}
F(x,\ha,\sigma) &\equiv F_{\rm N}(x) \, \hat{F}(x,\ha,\sigma)\, , \\
\label{eq:gi}
 F_N(x) &\equiv \dfrac{32}{5}\nu^2 x^5 \, ,
\end{align}
we obtain
\begin{align}
\label{eq:gi_NN}
\hat{F}(x,\ha,\sigma)&\equiv 1 - \dfrac{1247}{336}x + \left(4\pi - \dfrac{11}{4} \ha -\dfrac{5}{4}\sigma\right)x^{3/2}\nonumber\\
            &+ \left(-\dfrac{44711}{9072}+\dfrac{33}{16}\ha^2 + \dfrac{31}{8} \ha \sigma\right)x^2\nonumber\\
            &+ \left(-\dfrac{8191}{672} \pi-\dfrac{59}{16}\ha - \dfrac{13}{16}\sigma\right)x^{5/2}.
\end{align}
Note the minus sign in front of $(59/16)\ha$,
which corrects the wrong plus sign in Eq.~(5.19) of 
Tanaka {\it et al}.~\cite{Tanaka:1996ht}.
It is also pleasing to note that Eq.~(\ref{eq:gi_NN}) agrees, up to the expected PN-order,
with the respective PN-prediction for the comparable mass case,
Eq.~(4.9) of~\cite{Marsat:2013caa},
when restricted to the test-particle limit $\nu \rightarrow 0$.

For completeness, we also rewrite the multipolarly decomposed fluxes,
which were provided by Tanaka {\it et al}.\ in terms of $u$,
in terms of $x$.
It is important to mention that the multipolar decomposition
is performed here with respect to the spin-weighted
spheroidal harmonics, $S_{s\lm}$.
Each multipolar flux
can be separated into its LO contribution
and higher-order corrections,
$F_{ S_{\ell m}}\equiv F_{\ell m}^{\rm LO} \, \hat{F}_{S_{\ell m}}$.
In this way we obtain
\begin{widetext}
\begin{align}
\label{eq:22gi}
\hat{F}_{S_{22}} & = 1 -\dfrac{107}{21}x + \left(4\pi - \dfrac{8}{3}\ha - \dfrac{4}{3}\sigma\right)x^{3/2}+\left(\dfrac{4784}{1323}+2 \ha^2 + 4 \ha \sigma\right)x^2 
                        + \left(-\dfrac{428}{21}\pi + \dfrac{52}{27} \ha + \dfrac{208}{63}\sigma\right)x^{5/2},\\
\hat{F}_{S_{21}} & = 1 + 3(\sigma-\ha)x^{1/2}+\left(-\dfrac{17}{14}+\dfrac{9}{4}\ha^2-\dfrac{9}{2}\ha\sigma\right)x
                        + \left(2\pi + \dfrac{215}{252}\ha - \dfrac{367}{28}\sigma\right)x^{3/2},\\
\hat{F}_{S_{33}} & = 1 - 8x + (6\pi-3\sigma - 4 \ha)x^{3/2}, \\
 \label{eq:32gi}
\hat{F}_{S_{32}} & = 1 + \left(4\sigma -\dfrac{8}{3}\ha\right)x^{1/2},\\
\label{eq:31gi}
\hat{F}_{S_{31}} & = 1 - \dfrac{16}{3}x + \left(2\pi +5\sigma - \dfrac{100}{9}\ha\right)x^{3/2} \, .
\end{align}
\end{widetext}
In the main body of this paper we focus on the total flux,
but in Appendix~\ref{app:CompInfos} also these formulas are compared
(and found to be consistent) with the
corresponding numerical results.

\subsection{Comparing numerical and PN total GW fluxes}
\label{pnnum_compare}
\begin{figure}[!htbp]
  \centering  
  \includegraphics[width=0.48\textwidth]{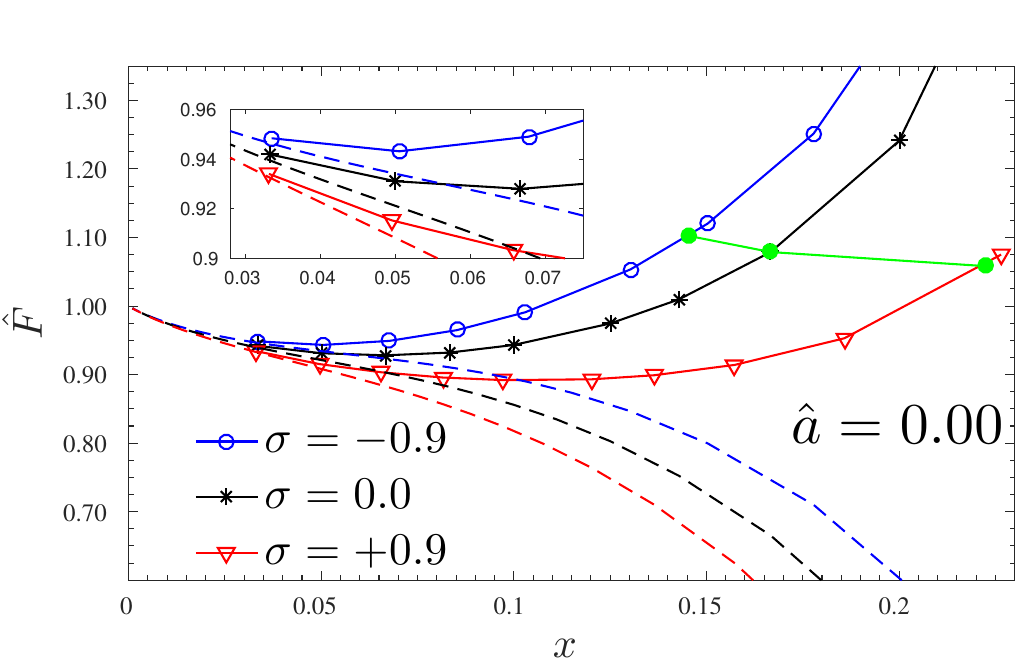} 
  \includegraphics[width=0.48\textwidth]{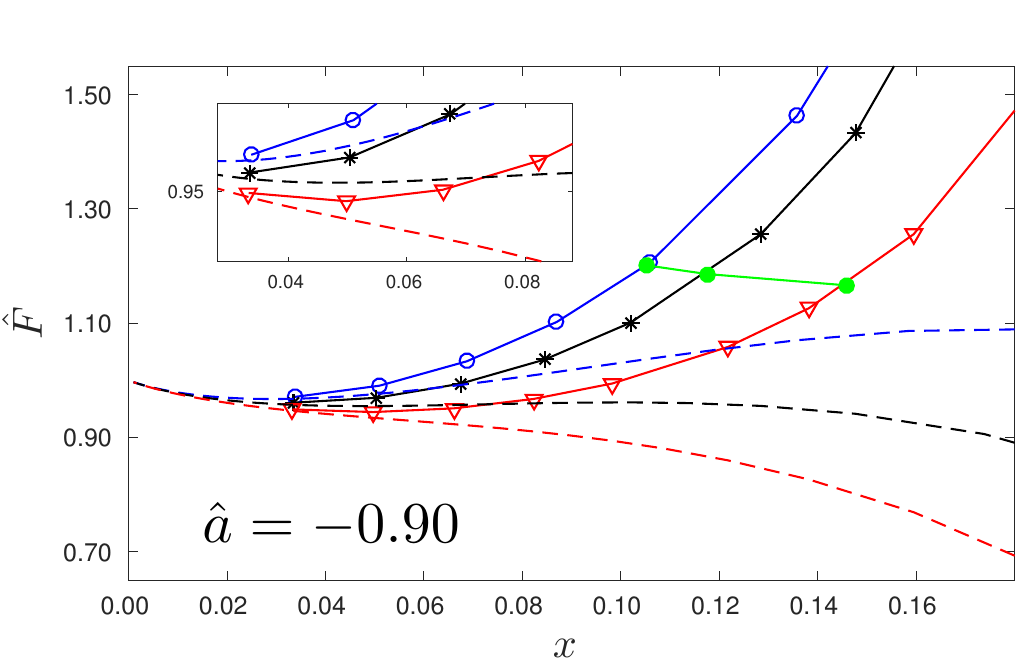} 
  \includegraphics[width=0.48\textwidth]{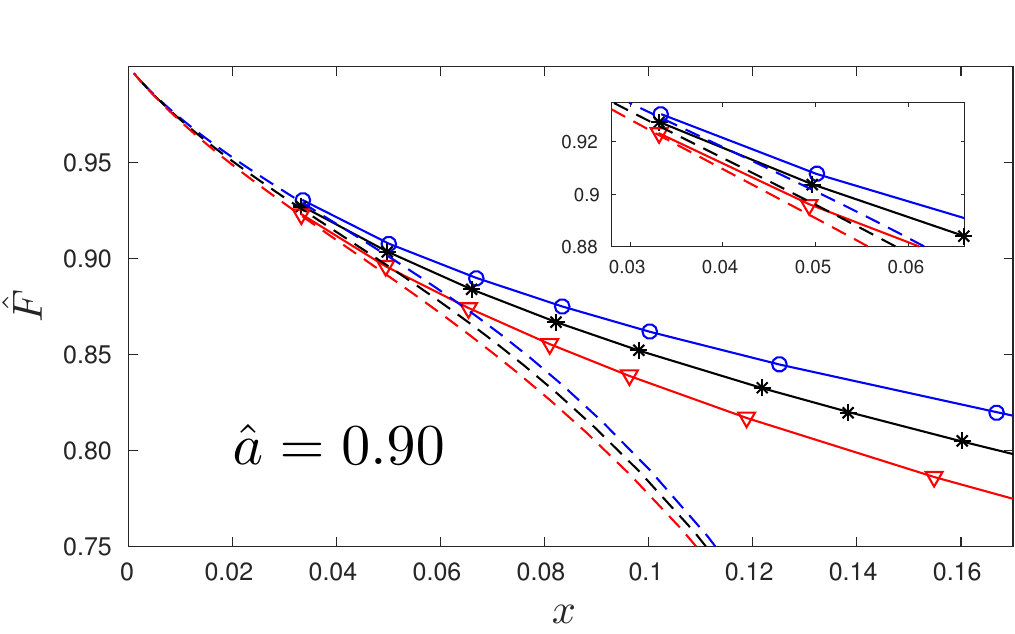} 
  \caption{
    Comparison of numerical (solid) and analytical PN (dashed)
    total energy fluxes over $x=\hat{\Omega}^{2/3}$.    
    We consider three values of the background spin,
    $\ha=0.0$ (top panel),
    $\ha=-0.9$ (middle panel), and
    $\ha=+0.9$ (bottom panel).
    For each value of $\ha$, we plot the fluxes
    for a spinning particle with spins    
    $\sigma=-0.9$ (blue, circles),
    $\sigma=0.0$ (black, stars) with data computed
    by Hughes~\cite{Hughes:1999bq,Hughes:2001jr,Sundararajan:2007jg,Taracchini:2013wfa},
    see text, and
    $\sigma=+0.9$ (red, triangles)
    over $x$.    
    The numerical fluxes are obtained summing up the modes with $m=1,2,3$;
    the PN fluxes
    are given by Eq.~\eqref{eq:gi_NN},
    which sums up only the modes considered in~\cite{Tanaka:1996ht}.
    The green filled circles mark
    the interpolated fluxes
    at the locations of the LSO, cf.\ Table~\ref{tab:LSOs}.
    At low orbital frequencies $\hat{\Omega} \to 0$,
    or equivalently $x\to0$ and $\hr \to \infty$,
    the numerics are consistent with the PN-prediction,
    reaching in all cases
    a fractional difference $ \leq 0.5\%$
    at our outermost data point $\hr=30$,
    see also 
    Tables~\ref{tab:full_infos_a00}-\ref{tab:full_infos_ap9}.
     }
  \label{fig:Ftot}
\end{figure}

Now, we compare
and contrast the outcome of our 
numerical computation of the full energy flux with the 
2.5PN information, as given by Eq.~\eqref{eq:gi_NN}.
This comparison is only meaningful in a finite region
of parameter space because our numerical results 
are inaccurate for too large radii ($r>30M$)
while the PN prediction is not expected to be reasonable
for too small radii (say, rather arbitrarily, $r\lesssim15M$).
At any rate, we have to settle with observing the
expected trend when traversing the region of common validity
$r\in[15M,30M]$
but are not able to cross-check numerics with analytics
at an arbitrary accuracy in the region $r\rightarrow \infty$..

The main result of our study is the unambiguous
trend of the total energy flux,
approximated as the sum over the $m=1,2,3$ contributions
in our numerical data,
against the 2.5PN prediction Eq.~\eqref{eq:gi},
as highlighted in Fig.~\ref{fig:Ftot}. 
To our knowledge, it is the 
first time that this comparison is done.
Figure~\ref{fig:Ftot} shows the numerical
fluxes together with the 
2.5PN predictions for several values
of $\ha$ and $\sigma$.
The $\sigma=0$ data (blue, circles)
in Fig.~\ref{fig:Ftot}
are not produced with our code, but
were computed by S.~Hughes
at extreme accuracy in the frequency domain
and were kindly made available to us
for completing the comparison here.
To be consistent with our $\sigma\neq0$ data, 
we compute the $\sigma=0.0$ total flux
by summing over multipoles $m=1,2,3$
and $\ell=1,..8$.
Note that we make the choice to
stay consistently at 2.5PN
though higher-order information
up to 22PN for Schwarzschild~\cite{Fujita:2012cm}
and 20PN for Kerr~\cite{Shah:2014tka}
are available in the literature for a
nonspinning particle.

Inspecting Fig.~\ref{fig:Ftot},
one can draw several conclusions
from our numerical computations.
First, 
if $\sigma <0$, the GW flux is increased with respect
to the $\sigma=0$ case, while,
if $\sigma > 0$, it is decreased. 
This holds independently of the background spin $\ha$,
i.e.\ for all three panels of Fig.~\ref{fig:Ftot}.
Second, one sees that the prediction of the approximate PN series
is qualitatively consistent with the numerical data
because there is a clear trend of numerics (solid lines)
towards analytics (dashed lines) as $x\to0$;
one sees that the disagreement between PN
and numerical data progressively
decreases as $x$ becomes smaller.
In particular, the spin-dependence is 
captured correctly in the numerics,
i.e.\ though evidently all the normalised
fluxes converge towards $\sim 1$, and thus
towards one another,
the offset between the $\sigma=0.0$ and the $\sigma\neq0$
lines is consistent in the numerics and
in the analytics,
as highlighted in the small insets in Fig.~\ref{fig:Ftot}.
Quantitatively, the visual comparison
is supported by the data shown in 
Tables~\ref{tab:full_infos_a00}-\ref{tab:full_infos_ap9}
in Appendix~\ref{app:Tables}.
For example, focusing first on the $\ha=0$ case, one sees that 
at $\hr=20$, i.e.\ $x\approx0.05$, the relative deviations
amount
to $1\%$ for $\sigma=-0.9$ and
to $0.7\%$ for $\sigma=0.9$.
At $\hr=30$, i.e.\ $x\approx0.033$,
they have decreased to
$0.23\%$ for $\sigma=-0.9$ and
$0.14\%$ for $\sigma=0.9$.
This is almost reaching the
level of accuracy, $\sim0.2\%$ of our numerical results,
as estimated in Sec.~\ref{sub:nums}.
A similar consistency is found when $\ha\neq 0$. 
For $\ha=-0.9$, 
at $\hr=30$ the agreement has reached
$0.33\%$ for $\sigma=-0.9$ and
$0.25\%$ for $\sigma=0.9$.
For $\ha=+0.90$, the fractional difference at $\hr=30$ 
between the numerical and the PN prediction is
$0.16\%$ for $\sigma=-0.9$ and
$0.08\%$ for $\sigma=0.9$.
These results mutually confirm
the 2.5PN predictions as well 
as our numerical computations.

Finally, we mention that the
consistency check for $x\to0$ can be repeated in the
same way for the multipolar fluxes, Eqs.~\eqref{eq:22gi}-\eqref{eq:31gi},
instead of the total flux.
Without illustration, we have included the corresponding
data in Table~\ref{tab:full_infos_a00}-\ref{tab:full_infos_ap9}.
The trend is again unambiguous, though
a bit worse than in the total flux, especially if $(\ell,m)\neq(2,2)$;
likely, simply because the subdominant modes 
themselves are,
(i)~at 2.5PN known with too few terms
after the LO-contribution; e.g., the
formula for the flux in the $(3,2)$-mode,
Eq.~\eqref{eq:32gi},
only includes one term beyond LO, and
(ii)~smaller in absolute value and, therefore,
relatively more affected by the numerical noise.
For more details,
see the multipolar analysis of the fluxes
with respect to the spin $\sigma$ at a fixed radius
in Appendix~\ref{app:CompInfos}
(notably, discussed there in a one-to-one comparison
with the formulas given by~\cite{Tanaka:1996ht},
i.e.\ without the change of variables
$u \to x$).

\section{Conclusions}
\label{sec:conc}

In this paper we have presented a new computation of
the gravitational wave fluxes emitted by 
a spinning particle on circular equatorial orbits
of a Kerr black hole.
This is done, for the 
first time, by solving the Teukolsky equation in the time domain,
using the {\tt Teukode} of~\cite{Harms:2014dqa, Nagar:2014kha},
for a pole-dipole particle source term that is
built according to the Mathisson-Papapetrou-Equations
under the Tulczyjew-spin-supplementary condition.
We investigated three values of the background spins,
$\ha=0$ and $\ha=\pm0.9$,
and ten values of the particle spin
$\sigma/ \nu \in \pm \{0.1,0.3,0.5,0.7,0.9\}$.
We stress that there is no technical obstruction
to obtain data at even higher rates for the spins.

About 20 years ago Tanaka {\it et al}.~\cite{Tanaka:1996ht}
computed the GW energy fluxes for the considered test-particle setup
at the 2.5PN order. Recently, Marsat {\it et al}.~\cite{Marsat:2013caa}
performed the respective comparable-mass PN calculation
(see also~\cite{Blanchet:2011zv, Blanchet:2012sm, Bohe:2013cla}),
which contains Tanaka {\it et al}.'s result as a special case
in the test-particle limit and extends it by terms
up to 4PN. While Tanaka {\it et al}.'s results were thus analytically
confirmed, a detailed numerical check of their accuracy was missing.
In this work, for the first time, we proved the consistency between
numerical results and the post-Newtonian
calculations of the GW fluxes at 2.5PN.
More precisely, at our
outermost data point, $\hr=30$,
we find a relative disagreement between the
total numerical flux, approximated as the sum over 
$m=1,2,3$ mode contributions, 
and the 2.5PN result, Eq.~(5.19) of~\cite{Tanaka:1996ht},
at the order of $\leq 0.5 \%$  
for all considered values of $\ha$ and $\sigma$.
This mutually confirms the numerically untested 2.5PN
prediction and our numerical computations.
Moreover, we mention that our results seem to disagree
quantitatively
with the corresponding numerical results of~\cite{Han:2010tp},
though we observe a certain qualitative consistency (see App.~\ref{app:CompInfos}).
We have presented our results in the form of
a central plot, Fig.~\ref{fig:Ftot},
and supported the visual impression quantitatively
by data tables, Tables~\ref{tab:full_infos_a00}-\ref{tab:full_infos_ap9}.
Notably, such database is missing from the literature
and will provide a valuable orientation for 
future studies in the same direction.
Additionally,
the numerical data presented here
will serve as a test bed for developing suitably
resummed expressions for the PN fluxes -
a procedure which was successfully developed
for a nonspinning particle
and which drastically improved the 
regime of accuracy of straight PN expressions 
towards the strong-field regime~\cite{Damour:2008gu,Pan:2010hz,Fujita:2014eta}.
Thus, our numerical results will have
immediate impact in modeling analytically the 
radiation reaction force for a spinning particle 
in quasi-circular, adiabatic inspiral motion
about a rotating BH background,
and, in turn, to the effective-one-body
model with generic spins.

\begin{acknowledgments}
We want to thank B.~Br{\"u}gmann, G.~Sch{\"a}fer, T.~Damour, 
D.~Bini, S.~Balmelli and D.~Hilditch for useful discussions
on the topic and helpful comments on the manuscript.
We are especially grateful to Scott Hughes for providing his data
on circular orbits for a nonspinning particle.
This work was supported in part by  DFG grant SFB/Transregio~7
``Gravitational Wave Astronomy''. 
E.H thanks IHES for hospitality during the development of part of this work.
G.L-G is supported by UNCE-204020a and by GACR-14-10625S.
\end{acknowledgments}

\appendix

\section{Complementary discussions}

\label{app:CompInfos}

In this Appendix we provide some
complementary information 
on the comparison of our numerical results
with the 2.5PN predictions of~\cite{Tanaka:1996ht}.
First, we perform the analysis multipole by multipole,
discussing the cases $\lm=22,21,33,32$.
This comparison
is conducted in a slight
alteration with respect to the main body of this paper
because here we remain with the variable $u\equiv M/r$
instead of switching to $x\equiv \hat{\Omega}^{2/3}$.
Note that this allows the check of the, so to say,
raw result of the computations
performed in~\cite{Tanaka:1996ht};
the switch $u \to x$, which we used in the main body,
is necessary to argue in terms
of usual PN parameters, but it amounts
to an additional approximation by using the PN expansion
of the frequency, Eq.~\eqref{eq:PN_Omega}, which,
in particular, is only linear in the spin.
In the second part, 
we will then repeat the consistency check
of the full energy flux in terms of $u$,
complementing the comparison shown in the
main body in terms of $x$.

\subsection{Spin-weighted spheroidal harmonics}
\label{sec:Sslm}

While the focus in the main body of the paper
is on the full flux, the data in 
Tables~\ref{tab:full_infos_a00}-\ref{tab:full_infos_ap9},
and the discussion in this appendix 
include also comparisons of the separate multipolar
contributions.
Therefore, we emphasize again that the results of Tanaka 
{\it et al}.~\cite{Tanaka:1996ht} on multipolar fluxes, as well 
as the rewriting of these results in terms of other 
variables, Eqs~\eqref{eq:22gi}-\eqref{eq:31gi},
refer to a decomposition with respect
to the spin-weighted spheroidal
harmonics, $S_{s\lm}$.

Spin-weighted spheroidal harmonics
arise in the traditional
separation of the original TE in the frequency domain.
The angular part of the resulting
ordinary differential equations, 
Eq.~(2.7) in \cite{Teukolsky:1974yv},
describes an eigenvalue problem for
each relevant $\omega$ of the solution.
The eigenfunctions are what we call
$S_{s\lm}$, with eigenvalue $E_{s\lm}$.
For $\ha=0$ they reduce to the well-known spin-weighted
spherical harmonics (see, e.g., \cite{Bruegmann:2011zj}).
In the case of circular equatorial orbits $\omega$
is fixed, and thus there is a unique set of $S_{s\lm}$
functions, which forms a complete basis.
The analytic formulas of Tanaka {\it et al}.\
are given with respect to a decomposition in
terms of the $S_{s\lm}$-basis.
Hence, to make comparisons of the multipolar fluxes 
we also project our full solution onto 
modes with respect to the $S_{s\lm}$-basis.
The computation of the $S_{s\lm}$ is not trivial,
but a nice procedure is given 
in Appendix A of~\cite{Hughes:1999bq},
which we adopted here
(cf.\ also ~\cite{Teukolsky:1973ha,Press:1973zz}).

\subsection{Multipolar analysis at at fixed radius}

Let us recall the 2.5PN results for the 
multipolar fluxes,
Eqs~(5.16) of~\cite{Tanaka:1996ht}. 
The stated quantities are
the normalized fluxes 
$\eta_{\lm} = F_{S_\lm} / \left( 32/5\, u^5 \right)$
for $\lm=\{22,21,33,31,32,44,42\}$,
where $F_{S_\lm}$ is the multipolar flux in the
$\ell\pm m$-modes,
with the definition of the $F_{S_\lm}$ used above,
see also Eq.~(2) of \cite{Damour:2008gu}.
It is important to note that
the normalisation factor is, in general,
not the usual Newtonian flux,
$F^N_{22}= (32/5)\, x^5$.
Thus one has to mind the different
normalisation with respect to that
used in the main body of this paper.
Concretely, the multipolar 2.5PN-formulas 
as given by~\cite{Tanaka:1996ht} versus $u$ read,
\begin{align}
 \label{eq:TanakaFluxes_22}
 & \eta_{22}(\ha,\sigma,u) 
 = 1 
  - \frac{107}{21} u
  + \left(4 \pi - 6 \ha - \frac{19}{3} \sigma \right) u^{1.5}
  \nonumber \\
 &+ \left(\frac{4784}{1323} + 2 \ha^2 + 9 \ha \sigma \right) u^2 
  \nonumber \\
 &+ \left(-\frac{428}{21}\pi + \frac{4216}{189} \ha + \frac{2134}{63}\sigma \right) u^{2.5}
 \;, \\ 
 \label{eq:TanakaFluxes_21}
 & \eta_{21}(\ha,\sigma,u) 
 = \frac{1}{36} u 
  + \left(-\frac{1}{12} \ha + \frac{1}{12} \sigma \right) u^{1.5} 
   \nonumber \\
 &+ \left( -\frac{17}{504} + \frac{1}{16} \ha^2 - \frac{1}{8} \ha \sigma \right) u^2
 \nonumber \\
 &+ \left( \frac{1}{18} \pi - \frac{793}{9072} \ha - \frac{535}{1008} \sigma \right) u^{2.5}
 \;,  \\
 \label{eq:TanakaFluxes_33}
 & \eta_{33}(\ha,\sigma,u) 
 = \frac{1215}{896} u 
  - \frac{1215}{112} u^2 \nonumber \\
 &+ \left( \frac{3645}{448} \pi - \frac{1215}{112} \ha - \frac{10935}{896} \sigma \right) u^{2.5}
 \;,  \\
 \label{eq:TanakaFluxes_32}
 & \eta_{32}(\ha,\sigma,u) 
 = \frac{5}{63} u^2 
 + \left(-\frac{40}{189} \ha 
 + \frac{20}{63} \sigma \right) u^{2.5}
 \;, \\
 \label{eq:TanakaFluxes_31}
 & \eta_{31}(\ha,\sigma,u) 
 = \frac{1}{8064} u
  - \frac{1}{1512} u^2 \nonumber \\
 &+ \left( \frac{1}{4032} \pi  - \frac{17}{9072} \ha - \frac{1}{8064} \sigma \right) u^{2.5}
  \quad .
\end{align}
Unfortunately, the $31$-mode is too weak to be measured accurately
in our numerics, especially for orbits in
the weak-field, which we rely on for the validation of our data.
Also, the $44$ and $42$ energy fluxes do not contain spin-dependence at 2.5PN.
Consequently, we restrict the comparison below to the $\lm=\{22,21,33,32\}$ modes.
For our comparison we find it convenient to
further normalize each $\eta_{\lm}$ to the corresponding LO term.
For clarity, reading off the LO terms from
Eqs~\eqref{eq:TanakaFluxes_22}-\eqref{eq:TanakaFluxes_31}
and combining with $F^N_{22}$, we obtain the following expressions:
$\hat{\eta}_{22} = \eta_{22} = F_{22} / \left( \frac{32}{5} u^5 \right)$,
$\hat{\eta}_{21} = F_{21} / \left( 8 u^{6}/45 \right)$,
$\hat{\eta}_{33} = F_{33} / \left( 243 u^{6}/28 \right)$,
$\hat{\eta}_{32} = F_{32} / \left( 32 u^{7}/63 \right)$,
which are all of the form $\sim 1+\mathcal{O}(u)$.

\begin{figure*}[!htbp]
  \centering  
  \includegraphics[width=0.48\textwidth]{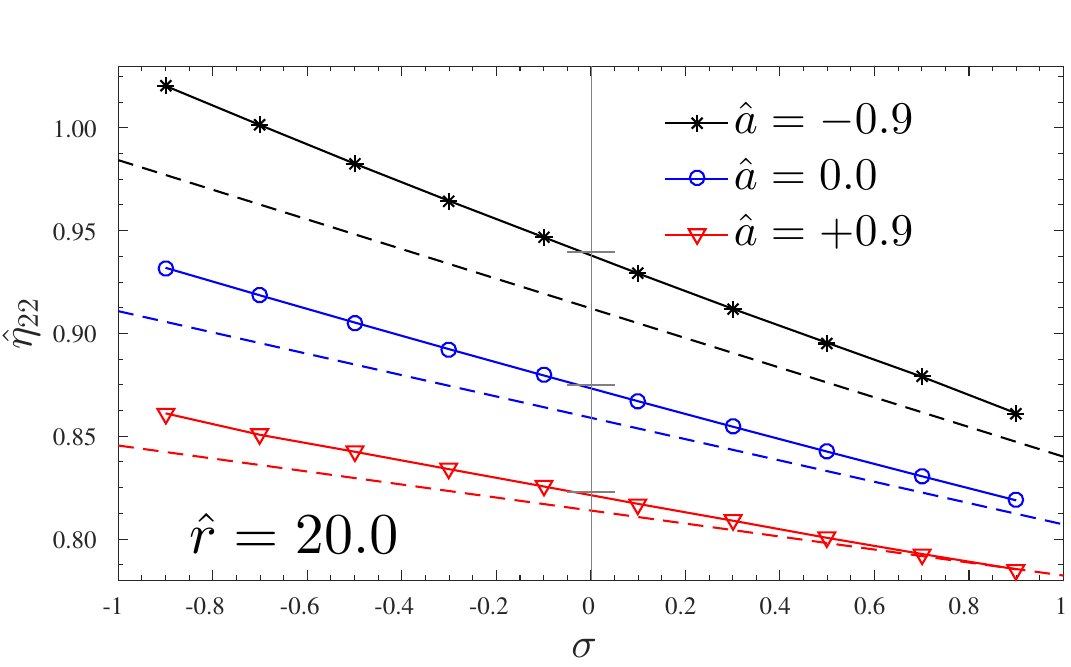} 
  \includegraphics[width=0.48\textwidth]{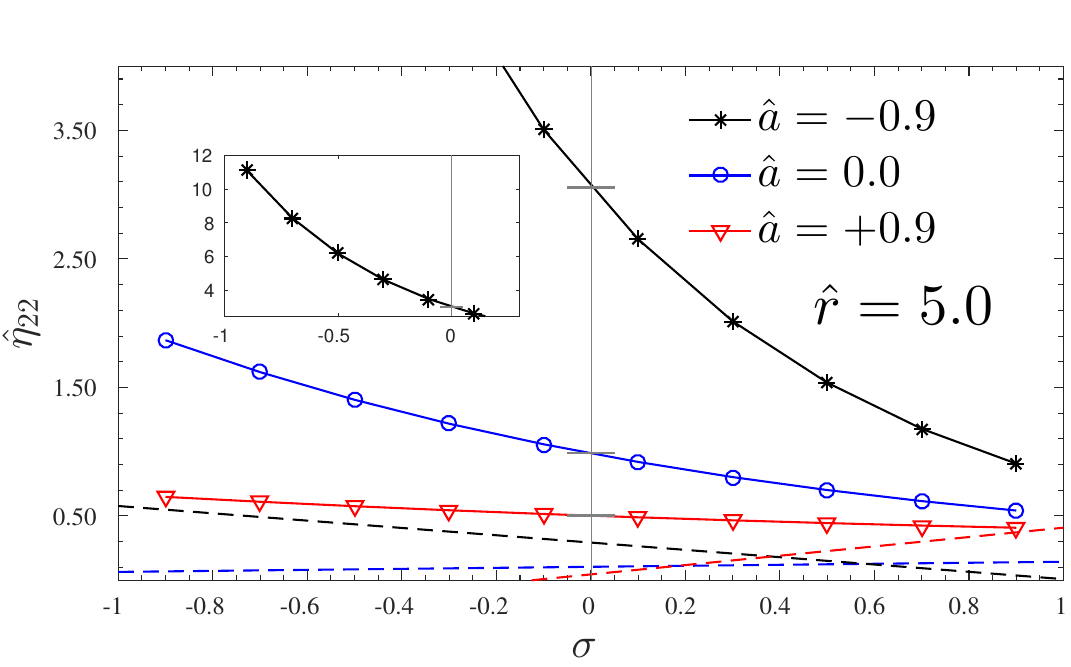} \\
  \includegraphics[width=0.48\textwidth]{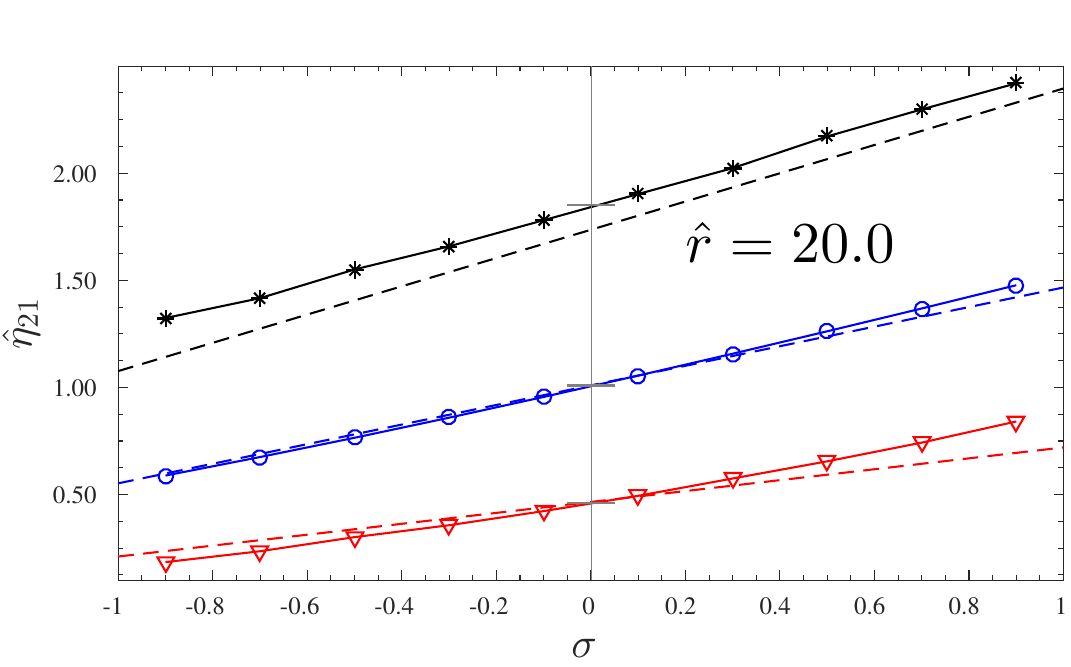} 
  \includegraphics[width=0.48\textwidth]{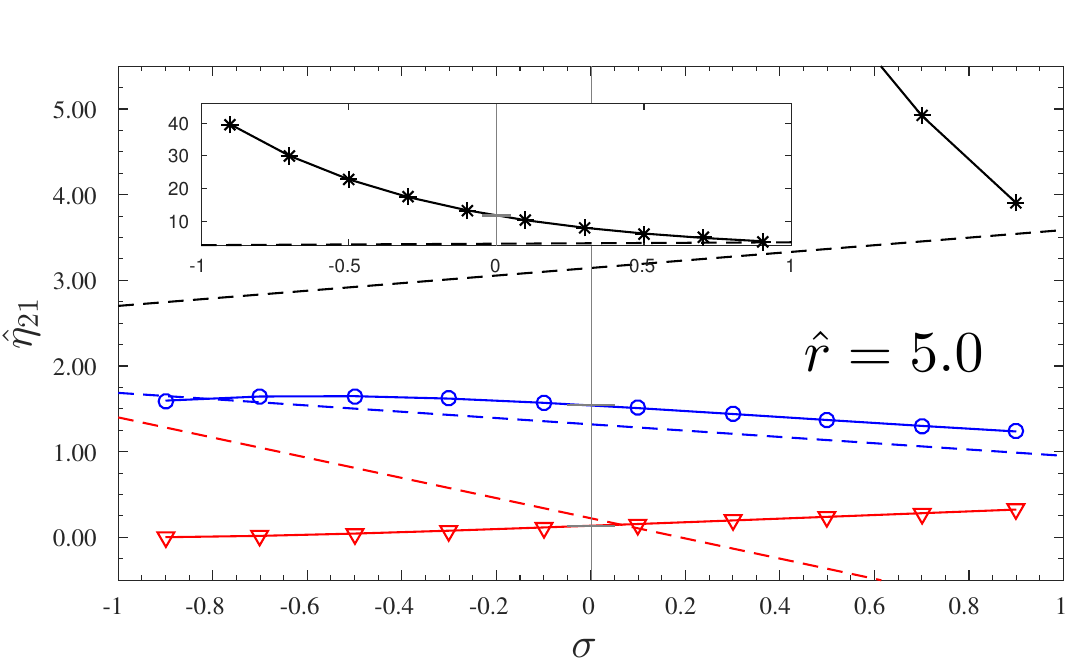} \\
  \includegraphics[width=0.48\textwidth]{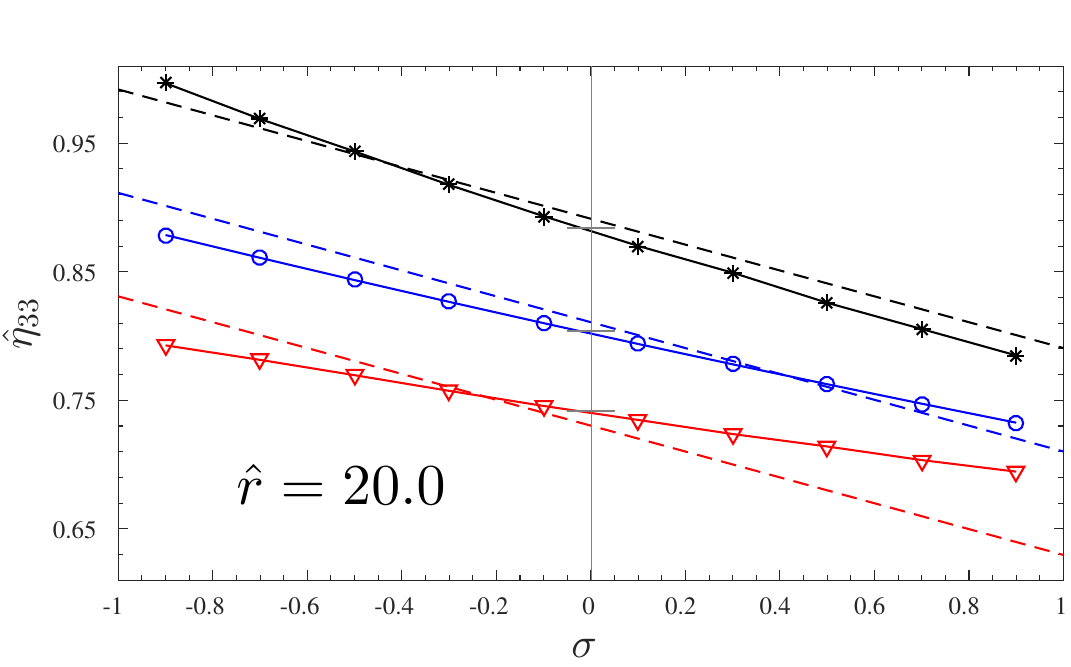}  
  \includegraphics[width=0.48\textwidth]{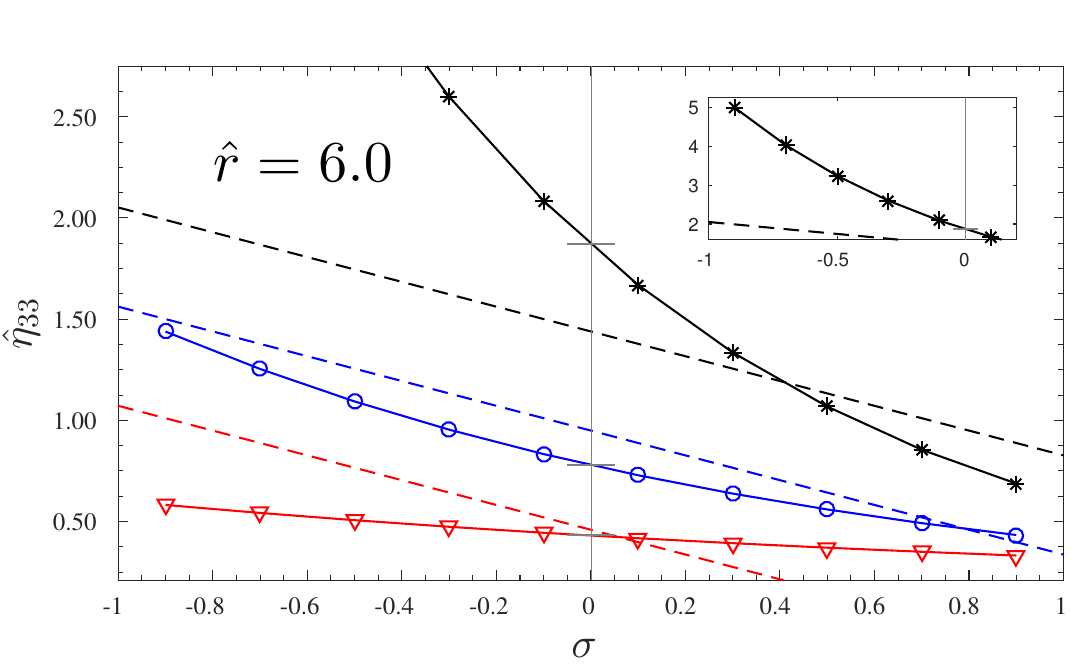} \\
  \includegraphics[width=0.48\textwidth]{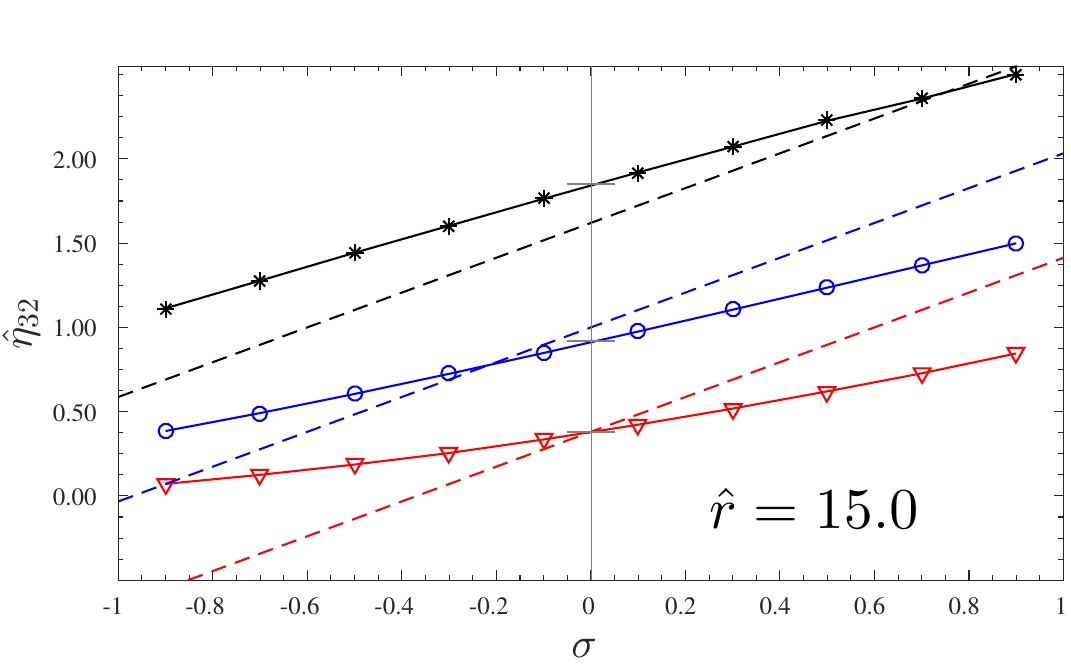}  
  \includegraphics[width=0.48\textwidth]{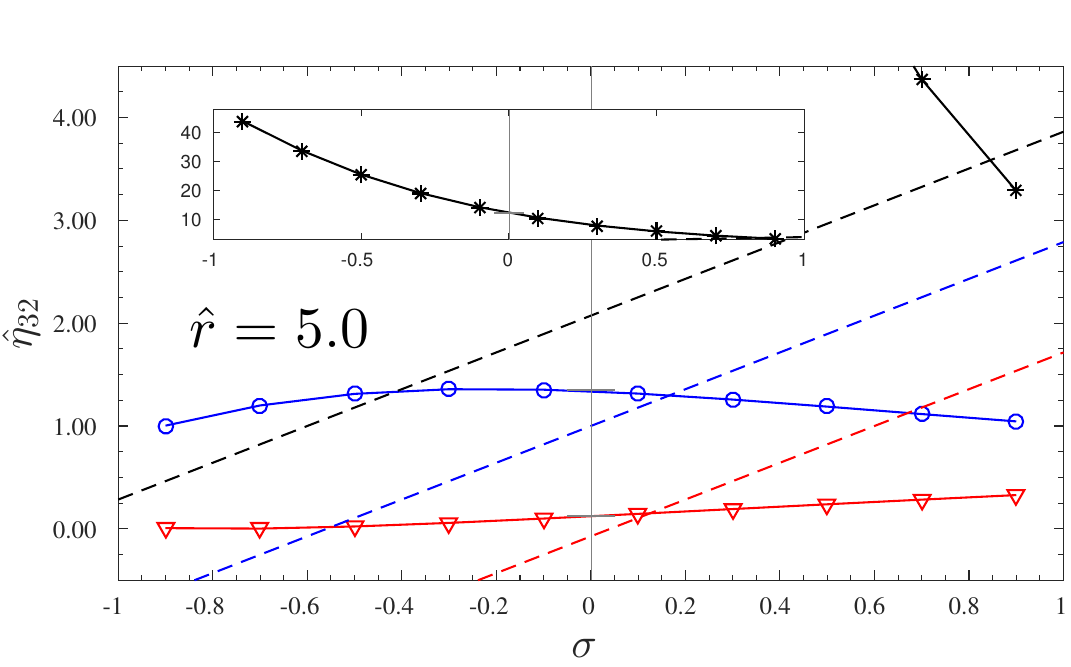} 
  \caption{
  Dependence of the LO-normalised multipolar fluxes, $\eta{\lm}$ on the spin $\sigma$,
  comparing numerical results (solid) with 2.5PN analytical predictions (dashed),
  at a weak-field orbit (left)
  and a strong-field orbit (right),
  for the three background spins
  $\ha=0.0$ (blue, circles),
  $\ha=-0.9$ (black, stars), and
  $\ha=+0.9$ (red, triangles).
  The panels show the multipolar energy fluxes in the 
  $S_{-222}$ (top),
  $S_{-221}$ (top middle),
  $S_{-233}$ (bottom middle) and
  $S_{-233}$ (bottom)
  modes.
  The small horizontal lines on the $\sigma=0.0$ 
  axis refer to the results
  for a nonspinning particle from the frequency-domain data of Hughes~\cite{Hughes:1999bq}.
  Note, for clarity, that these plots show the $\hat{\eta}_\lm$,
  as obtained from Eqs.~\eqref{eq:TanakaFluxes_22}-\eqref{eq:TanakaFluxes_31}
  by normalizing with the LO-Newtonian expressions in terms of 
  $u\equiv 1/r$ (e.g., for $\lm=22$ with $32/5 \, u^5$), whereas the values stated in 
  Tables~\ref{tab:full_infos_a00}-\ref{tab:full_infos_ap9}
  refer to the $\hat{F}_{S_\lm}$ and are thus
  normalized  to the LO-Newtonian expressions in terms of 
  $x\equiv\hat{\Omega}^{2/3}$ (e.g., for $\lm=22$ with $32/5 \, x^5 $).
     }
  \label{fig:Eflux_scr_Slm_NNxr_overS}
\end{figure*}

In Fig.~\ref{fig:Eflux_scr_Slm_NNxr_overS} we have collected
at a glance our numerical results
for the multipolar fluxes $\hat{\eta}_\lm$ 
and contrasted them with the
2.5PN predictions. 
To guide the reader
through the many panels,
note first that
the different modes
$(\ell,m) \in \{ (2,2),(2,1),(3,3),(3,2) \}$
are arranged in Fig.~\ref{fig:Eflux_scr_Slm_NNxr_overS}
from top to bottom, where for each mode
we have one weak-field (left panels)
and one strong-field (right panels) comparison
of numericals fluxes with PN predictions.
For each comparison we consider the three
different background spins
$\ha=-0.9$ (black, stars),
$\ha=0.0$ (blue, circles),
and
$\ha=0.9$ (red, triangles).
The $\sigma$ data points are
computed at $\pm\{ 0.1,0.3,0.5,0.7,0.9\}$.
To illustrate the smooth connection of our data to 
the correct $\sigma=0.0$ values,
which are again made available to us
at highest accuracy by
Hughes~\cite{Hughes:1999bq,Hughes:2001jr,Sundararajan:2007jg,Taracchini:2013wfa},
we draw short, gray, horizontal lines at the
respective $\sigma=0$ values.

Let us discuss first the weak-field comparison
illustrated in the left column of panels.
For all four modes,
our numerical fluxes (solid lines) are
consistent with the 2.5PN predictions (dashed lines),
i.e.\ with the LO-normalised versions of 
Eqs.~\eqref{eq:TanakaFluxes_22}-\eqref{eq:TanakaFluxes_31}.
Only in the $(3,2)$-mode the spin-behaviour
is poorly captured by the analytics, but note that
in this case we show the comparison at an orbital
separation of $\hr=15$.
At $\hr=20$ and for $\sigma=-0.9$ the $(3,2)$-flux
was too small in absolute value to be measured
reliably. But even at $\hr=20$ we would not
expect a much better agreement because
the 2.5PN formula for the $(3,2)$-mode
reaches only one term beyond the leading order.
The comparisons between numerics and PN predictions 
for strong-field orbits are shown in the 
right column of panels.
It is apparent from the plot that there is a clear failure
of the PN formulas in the strong-field regime.
Especially for $\ha=-0.9$, in which case the 
LSO is located at $r_{\rm{LSO}}\approx6.6$ 
for $\sigma=+0.9$ and  
$r_{\rm{LSO}}\approx10.0$ for $\sigma=-0.9$
the numerical results are
completely off the prediction.
For $\ha=0.0$ (blue circles)
at least the fluxes in the
$(2,1)$-mode (top middle panel)
and in the $(3,3)$-mode (bottom middle panel)
are reasonable. 

Another useful information that one reads
off Fig.~\ref{fig:Eflux_scr_Slm_NNxr_overS}
regards the spin-dependence of the multipolar
fluxes at fixed radius. One finds that, for
each value of $\ha$ and for each multipole,
each dataset in Fig.~\ref{fig:Eflux_scr_Slm_NNxr_overS}
can be accurately fitted with a polynomial in
$\sigma$. More precisely, for weak-field orbits 
the $\sigma$-dependence is essentially linear.
By contrast, for strong-field orbits the polynomial 
can be of higher order, at most third order.
This is interpreted as a direct 
reflection of the Tulczyjew-SSC,
Eq.~\eqref{eq:v_p_TUL}, and the 
related $\sigma$-squared dependence
of the dynamics, in connection with the fact 
that the TE source term is linear in the spin.
The appearance of the Riemann-tensor in connection
with the $\mathcal{O}(\sigma^2)$ terms in 
Eq.~\eqref{eq:v_p_TUL} explains
the transition to completely linear
spin dependence in the weak-field.

As a final comment we mention
that one cannot directly
contrast these plots
with the numbers stated in 
Tables~\ref{tab:full_infos_a00}-\ref{tab:full_infos_ap9}
because the latter refer to 
the PN-formulas recast in
terms of $x$. 
While it is possible to
compute all the relevant 
quantities,
i.e.\ $F_{S_\lm}$, $\hat{F}_{S_\lm}$,
${\eta}_\lm$ and $\hat{\eta}_\lm$,
from one another
in our numerical data
the analytical predictions are
not strictly equivalent
because of the approximation of $\Omega$,
see discussion in Sec.~\ref{sub:pn}.
Though the general trend is the same,
the values of fractional differences
shown in Tables~\ref{tab:full_infos_a00}-\ref{tab:full_infos_ap9}
are not fully compatible with what
one would obtain from the
data corresponding to Fig.~\ref{fig:Eflux_scr_Slm_NNxr_overS}.
\subsection{Total flux with respect to $u$}

\begin{figure}[!htbp]
  \centering  
  \includegraphics[width=0.48\textwidth]{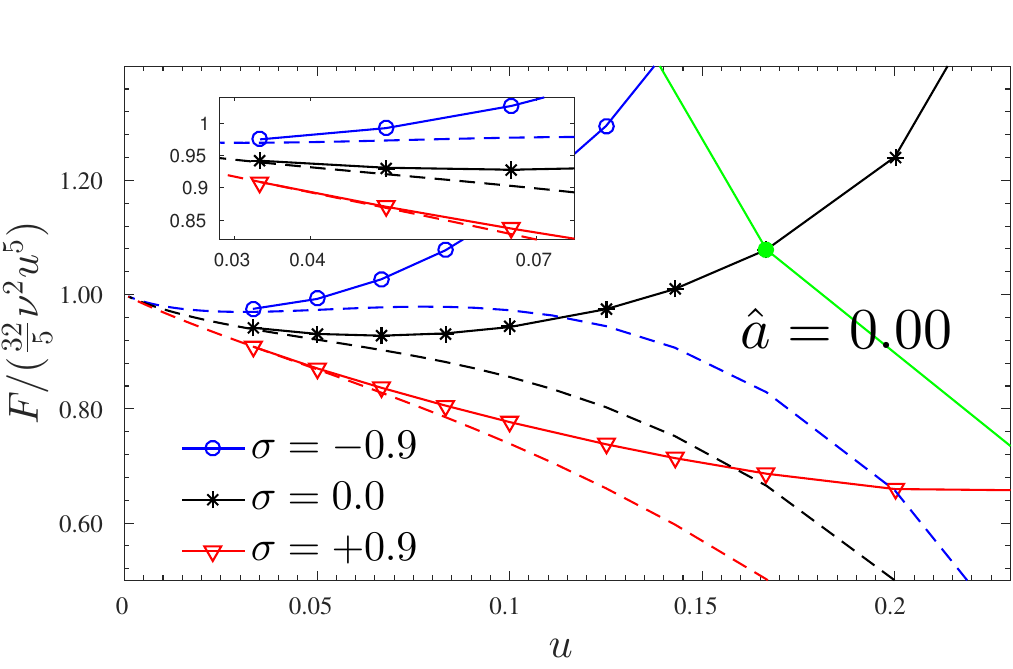} 
  \includegraphics[width=0.48\textwidth]{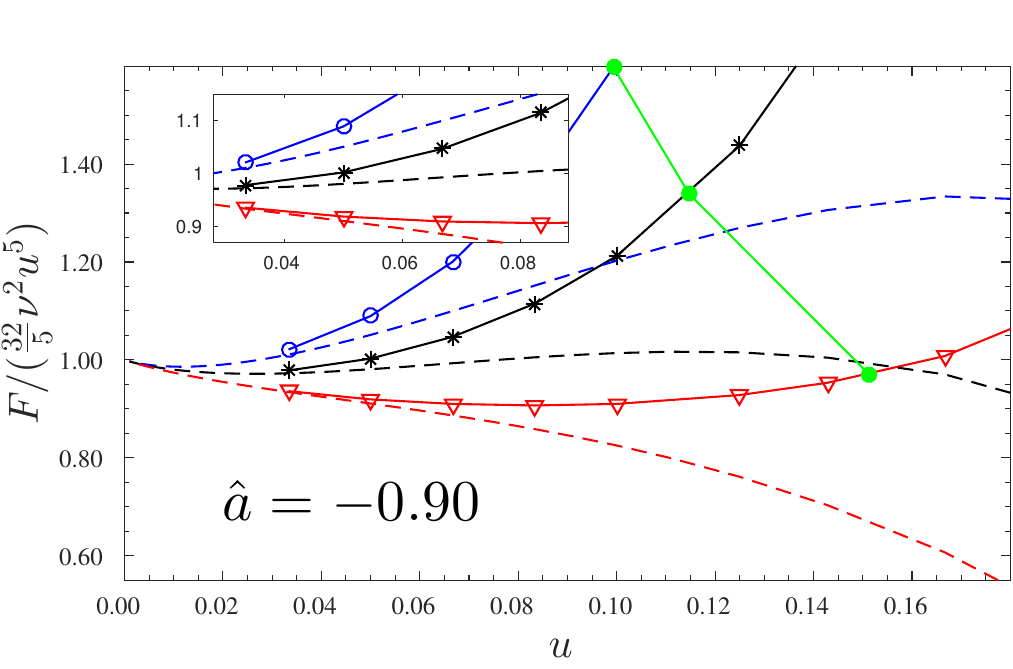} 
  \includegraphics[width=0.48\textwidth]{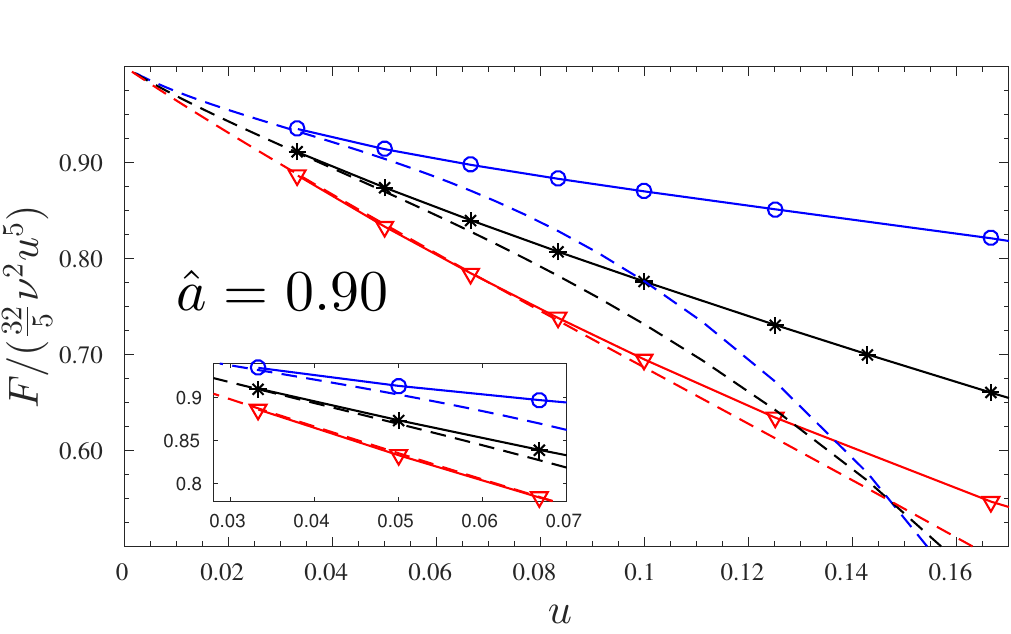} 
  \caption{ Comparison of numerical (solid) and analytical PN (dashed)
    total energy fluxes over $u \equiv M/r$.    
    Compare with Fig.~\ref{fig:Ftot} for further
    descriptions.
    In all three panels, i.e.\ for all considered background
    spins $\ha=0.0$ (top panel), $\ha=-0.9$ (middle panel), 
    and $\ha=+0.9$ (bottom panel), 
    we observe an unambiguous trend of our numerical
    data, approximated as the sum of $m=1,2,3$-modes,
    towards the 2.5PN prediction of Tanaka {\it et al}.~\cite{Tanaka:1996ht}
    as $u\to 0$.
     }
  \label{fig:Ftot_xr}
\end{figure}

For completeness, we repeat
the consistency check 
for the total energy flux
as obtained from our numerical
results and the 2.5PN predictions
in terms of $u\equiv M/r$. 
This means we normalize our full flux
here by $32/5 \, u^5$ instead of using
the conventional Newtonian flux $32/5 \, x^5$.
The relevant 2.5PN prediction of~\cite{Tanaka:1996ht}
is Eq.~\eqref{eq:F_bl} above.

Fig.~\ref{fig:Ftot_xr} compares the fluxes
for the three background spins $\ha=0.0$ (top panel), $\ha=-0.9$ (middle panel), 
and $\ha=+0.9$ (bottom panel) and
the three values of the particle spin
$\sigma=-0.9$ (blue, circles),
$\sigma=0.0$ (black, stars),
$\sigma=0.9$ (red, triangles).
In all cases there is a clear trend
of the numerical results (solid lines) towards
the analytical predictions (dashed lines).
In fact, comparing with Fig.~\ref{fig:Ftot}
the agreement seems to be even
more pronounced in this comparison over $u$.
This must be mainly attributed to the pure visual 
effect that originates from the larger
separation of the dashed lines from one another
in this representation.
On the contrary
computing the fractional differences between 
our numerics and the PN formula, Eq.~\eqref{eq:F_bl},
we find rather larger than smaller
differences compared to 
Tables~\ref{tab:full_infos_a00}-\ref{tab:full_infos_ap9}.
For example, 
at $\ha=-0.9$ and $\hr=20$ 
we find numerically for $\sigma=-0.9$ that
$F_{m\leq3} / ( \frac{32}{5}u^5)=1.089$, which is
$\sim 3.5\%$ off the PN prediction $ 1.0510$,
and for $\sigma=+0.9$ 
we find 
$F_{m\leq3} / ( \frac{32}{5}u^5)=0.9193$, which is
$\sim 0.91\%$ off the PN prediction $0.9109$.
Overall, this complementary comparison
of the total energy flux confirms
as well, visually even more pronouncedly,
a clear convergence of the numerical
results towards the PN predictions.

\subsection{Cross-check against [Phys.\ Rev.\ D 82, 084013 (2010)]}

Apart from the 2.5PN analytical prediction,
we can compare our numerical results for the
flux in the $S_{-222}$-mode with
the numerical results of~\cite{Han:2010tp},
in which the multipolar energy fluxes
from a spinning particle on a CEO are computed by
solving the TE in the frequency domain.
In that study, the dynamics are obtained 
within the same approximations that 
we use, i.e.\ solving the MPEQs with the TUL-SSC.
As checked in Sec.~\ref{sec:dynamics},
our dynamics agree at least in the orbital
frequency with that of~\cite{Han:2010tp},
which is a convincing indication that
we are, in fact, considering the same physical system.

\begin{figure}[!htbp]
  \centering  
  \includegraphics[width=0.48\textwidth]{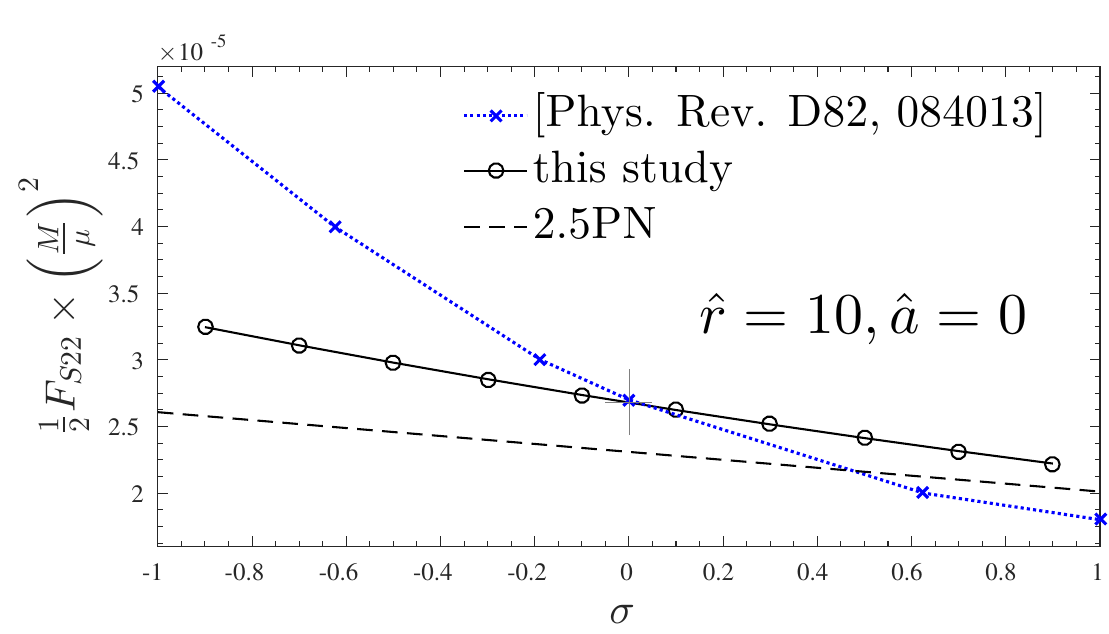} 
  \caption{
  Cross-check of the energy flux in the 
  $S_{-222}$-=mode as obtained in this study (black solid, circles)
  against the results of~\cite{Han:2010tp} (blue dotted, crosses),
  at $\hr=10$ and for $\ha=0.0$.
  The 2.5PN prediction is included for completeness (black dashed).
  The data marked by crosses were extracted visually
  by inspecting the top right panel in Fig.~4 in~\cite{Han:2010tp}.
  Note that here we show the \textit{not} Newton-normalised fluxes.
  The datapoints of~\cite{Han:2010tp} disagree qualitatively with 
  both our numerical data and the 2.5PN accurate analytical flux.
  Note that we have indications that the underlying dynamics is
  the same in both cases.
  Both numerical solutions are consistent with one another 
  at $\sigma=0.0$ and also with the 
  frequency-domain data of~\cite{Hughes:1999bq},
  which is marked by the small gray cross.
     }
  \label{fig:Eflux_WENBIAO}
\end{figure}

We can make a direct comparison for
$\hr=10$, $\ha=0$, in which case the 
({\it not} Newton-normalized) $S_{-222}$-flux
we considered was shown versus
the particle spin $\sigma$ in Fig.~4 of~\cite{Han:2010tp}. 
Our Fig.~\ref{fig:Eflux_WENBIAO}
draws a visual comparison between the results as: 
(i)~obtained by~\cite{Han:2010tp}, represented
as crosses joined by a dotted line,
(ii)~as computed with our numerical code;
(iii)~as obtained from the 2.5~PN approximate formula 
$F_{S_{22}}/\nu^2= 32/5 u^5\hat{\eta}_{22}$.
Note that in~\cite{Han:2010tp}
the $m=\pm2$ contributions were not summed together so that we need to 
compare it with $F_{S_{22}}/2$. Moreover, the data for the
crosses joined by the dotted line was read off
directly from Fig.~4 in~\cite{Han:2010tp}. The errors made in extracting the points 
from the plot are expected to be small in comparison to the differences we discuss 
below.
Our Fig.~\ref{fig:Eflux_WENBIAO} shows that the results of~\cite{Han:2010tp}
(blue dotted, crosses) are in {\it quantitative} disagreement
with both the post-Newtonian formula (black, dashed)
and our numerical computations (black solid, circles).
The reasons for the disagreement
are unclear at the moment.
Nevertheless, we are convinced of the
correctness of our implementation
because of the consistency with the 2.5PN expressions,
which was not shown in~\cite{Han:2010tp}.
Furthermore, note that the nonlinear shape of the 
blue dotted line with crosses
(the numerical results of~\cite{Han:2010tp})
resembles qualitatively what we find
for the $22$-mode at smaller radii;
e.g., see the top right panel of Fig.~\ref{fig:Eflux_scr_Slm_NNxr_overS},
whose blue line shows our results for $\ha=0.0$
at $\hr=5$ and is nonlinear in a similar way.

 \clearpage
 \onecolumngrid

\setlength{\headsep}{0.0in}

\section{Data tables} \vspace{-0.5cm}

\label{app:Tables}


\begin{table*}[!htbp]
\captionof{table}{Case $\ha =0$: fractional differences between numerical 
   and PN energy fluxes $1-F^{\rm PN}/F^{\rm Num}$. From left to right,
  the columns report: the dimensionless Boyer-Lindquist orbital
  radius $\hat{r}$; the dimensionless orbital frequency $\hat{\Omega}$;
  the PN-parameter $x=\hat{\Omega}^{2/3}$; the, Newton-normalized, total 
  numerical flux $\hat{F}_{(m\leq 3)}^{\rm Num}$ summed over $m=1,2,3$; the 
  corresponding fractional difference with the 2.5~PN result, Eq.~\eqref{eq:gi}
  the $\ell=m=2$ numerical contribution; $\hat{F}_{22}^{\rm Num}$; the
  fractional difference with the corresponding PN result, Eq.~\eqref{eq:22gi}; 
  and the analogous columns for $\ell=2,m=1$
  and $\ell=m=3$.}
\label{tab:full_infos_a00} 
\centering
\begin{tabular}[t]{| c | c c c | c c | c c | c c | c c |}
 \hline 
 \multicolumn{12}{|l|}{ {\bf\large{$\hat{a}=0.00$}} } \\ \hline 
 $\hat{r}$ & $\sigma$ & $M\Omega$ & $x$ & 
 $\hat{F}_{m\leq3}$ & $\Delta \hat{F}_{m\leq3} [\%]$  & 
  $\hat{F}_{\rm{S22}}$ & $\Delta \hat{F}_{\rm{S22}} [\%] $ & 
 $\hat{F}_{\rm{S21}}$ & $\Delta \hat{F}_{\rm{S21}} [\%] $ & 
 $\hat{F}_{\rm{S33}}$ & $\Delta \hat{F}_{\rm{S33}} [\%] $ \\ 
 \hline 
4.00  & -0.90 & 0.150524 & 0.282968 & 3.1651 & 99.49 & 2.2085 & 142.28 & 2.1513 & 9.75 & 2.4164 & 18.06 \\ 
      & -0.50 & 0.138047 & 0.267107 & 2.4858 & 96.74 & 1.7912 & 138.97 & 2.2948 & 27.12 & 1.8241 & 8.32 \\ 
      & 0.00 & 0.125000 & 0.250000 & 1.8220 & 95.55 & 1.3624 & 135.08 & 2.2378 & 33.78 & 1.2637 & 7.32 \\ 
      & 0.50 & 0.114518 & 0.235821 & 1.3433 & 87.11 & 1.0396 & 131.00 & 2.0934 & 32.59 & 0.8677 & 26.81 \\ 
      & 0.90 & 0.107796 & 0.226500 & 1.0750 & 82.60 & 0.8516 & 127.49 & 2.0062 & 29.44 & 0.6524 & 42.38 \\ 
 \hline 
5.00  & -0.90 & 0.101836 & 0.218072 & 1.5795 & 67.31 & 1.2118 & 102.08 & 0.9497 & 38.51 & 1.2152 & 19.31 \\ 
      & -0.50 & 0.095924 & 0.209549 & 1.4241 & 64.21 & 1.1119 & 96.38 & 1.2456 & 3.58 & 1.0572 & 20.66 \\ 
      & 0.00 & 0.089443 & 0.200000 & 1.2409 & 58.93 & 0.9885 & 89.46 & 1.5479 & 14.78 & 0.8771 & 23.82 \\ 
      & 0.50 & 0.083900 & 0.191650 & 1.0702 & 54.46 & 0.8683 & 82.53 & 1.7693 & 20.80 & 0.7157 & 28.88 \\ 
      & 0.90 & 0.080095 & 0.185810 & 0.9530 & 50.07 & 0.7831 & 77.04 & 1.9220 & 22.13 & 0.6088 & 32.55 \\ 
 \hline 
6.00  & -0.90 & 0.074985 & 0.177821 & 1.2515 & 43.27 & 1.0117 & 65.68 & 0.7009 & 42.85 & 0.9752 & 22.37 \\ 
      & -0.50 & 0.071724 & 0.172628 & 1.1731 & 41.23 & 0.9590 & 62.03 & 0.9790 & 11.12 & 0.8854 & 21.81 \\ 
      & 0.00 & 0.068041 & 0.166667 & 1.0788 & 36.08 & 0.8927 & 57.41 & 1.3231 & 7.40 & 0.7814 & 21.48 \\ 
      & 0.50 & 0.064772 & 0.161284 & 0.9836 & 34.69 & 0.8231 & 52.45 & 1.6346 & 15.02 & 0.6815 & 22.30 \\ 
      & 0.90 & 0.062438 & 0.157386 & 0.9142 & 31.66 & 0.7710 & 48.46 & 1.8745 & 18.07 & 0.6112 & 22.59 \\ 
 \hline 
7.00  & -0.90 & 0.058275 & 0.150310 & 1.1203 & 28.62 & 0.9369 & 43.31 & 0.6051 & 36.22 & 0.8833 & 19.24 \\ 
      & -0.50 & 0.056286 & 0.146871 & 1.0700 & 27.30 & 0.9016 & 41.05 & 0.8663 & 11.89 & 0.8211 & 18.19 \\ 
      & 0.00 & 0.053995 & 0.142857 & 1.0097 & 22.96 & 0.8577 & 38.14 & 1.2148 & 4.04 & 0.7488 & 16.85 \\ 
      & 0.50 & 0.051910 & 0.139156 & 0.9462 & 23.02 & 0.8097 & 34.78 & 1.5577 & 11.63 & 0.6765 & 16.39 \\ 
      & 0.90 & 0.050385 & 0.136416 & 0.8990 & 21.00 & 0.7733 & 32.07 & 1.8364 & 15.39 & 0.6244 & 15.70 \\ 
 \hline 
8.00  & -0.90 & 0.047019 & 0.130271 & 1.0534 & 19.64 & 0.9024 & 29.69 & 0.5603 & 28.04 & 0.8406 & 15.52 \\ 
      & -0.50 & 0.045716 & 0.127854 & 1.0174 & 18.76 & 0.8760 & 28.23 & 0.8085 & 10.73 & 0.7932 & 14.41 \\ 
      & 0.00 & 0.044194 & 0.125000 & 0.9745 & 15.18 & 0.8438 & 26.38 & 1.1527 & 2.33 & 0.7382 & 12.84 \\ 
      & 0.50 & 0.042785 & 0.122329 & 0.9279 & 15.87 & 0.8076 & 24.04 & 1.5063 & 9.41 & 0.6816 & 12.05 \\ 
      & 0.90 & 0.041737 & 0.120322 & 0.8931 & 14.50 & 0.7801 & 22.18 & 1.8029 & 13.45 & 0.6403 & 11.11 \\ 
 \hline 
10.00  & -0.90 & 0.033041 & 0.102968 & 0.9909 & 10.17 & 0.8762 & 15.47 & 0.5266 & 15.06 & 0.8087 & 9.84 \\ 
      & -0.50 & 0.032394 & 0.101619 & 0.9688 & 9.73 & 0.8591 & 14.79 & 0.7566 & 7.63 & 0.7769 & 8.93 \\ 
      & 0.00 & 0.031623 & 0.100000 & 0.9432 & 7.28 & 0.8389 & 13.99 & 1.0865 & 0.85 & 0.7405 & 7.51 \\ 
      & 0.50 & 0.030893 & 0.098456 & 0.9137 & 8.28 & 0.8149 & 12.72 & 1.4394 & 6.72 & 0.7009 & 6.77 \\ 
      & 0.90 & 0.030339 & 0.097275 & 0.8919 & 7.58 & 0.7970 & 11.77 & 1.7453 & 10.81 & 0.6720 & 5.92 \\ 
 \hline 
12.00  & -0.90 & 0.024867 & 0.085196 & 0.9648 & 5.79 & 0.8706 & 8.97 & 0.5210 & 7.12 & 0.8028 & 6.42 \\ 
      & -0.50 & 0.024499 & 0.084354 & 0.9494 & 5.55 & 0.8582 & 8.60 & 0.7380 & 5.20 & 0.7791 & 5.72 \\ 
      & 0.00 & 0.024056 & 0.083333 & 0.9321 & 3.79 & 0.8440 & 8.24 & 1.0533 & 0.32 & 0.7524 & 4.56 \\ 
      & 0.50 & 0.023631 & 0.082349 & 0.9110 & 4.73 & 0.8263 & 7.45 & 1.3959 & 5.15 & 0.7220 & 4.04 \\ 
      & 0.90 & 0.023304 & 0.081587 & 0.8957 & 4.34 & 0.8133 & 6.90 & 1.6974 & 9.06 & 0.7000 & 3.37 \\ 
 \hline 
15.00  & -0.90 & 0.017624 & 0.067724 & 0.9492 & 2.81 & 0.8738 & 4.53 & 0.5290 & 0.91 & 0.8085 & 3.64 \\ 
      & -0.50 & 0.017439 & 0.067248 & 0.9390 & 2.68 & 0.8651 & 4.36 & 0.7318 & 2.93 & 0.7917 & 3.18 \\ 
      & 0.00 & 0.017213 & 0.066667 & 0.9281 & 1.55 & 0.8559 & 4.29 & 1.0276 & 0.04 & 0.7733 & 2.31 \\ 
      & 0.50 & 0.016994 & 0.066100 & 0.9135 & 2.28 & 0.8432 & 3.79 & 1.3515 & 3.75 & 0.7507 & 2.04 \\ 
      & 0.90 & 0.016824 & 0.065658 & 0.9033 & 2.09 & 0.8342 & 3.52 & 1.6394 & 7.32 & 0.7347 & 1.59 \\ 
 \hline 
20.00  & -0.90 & 0.011352 & 0.050511 & 0.9433 & 1.00 & 0.8858 & 1.81 & 0.5537 & 3.00 & 0.8266 & 1.69 \\ 
      & -0.50 & 0.011275 & 0.050282 & 0.9370 & 0.94 & 0.8802 & 1.74 & 0.7396 & 1.04 & 0.8156 & 1.43 \\ 
      & 0.00 & 0.011180 & 0.050000 & 0.9312 & 0.32 & 0.8750 & 1.83 & 1.0090 & 0.06 & 0.8042 & 0.81 \\ 
      & 0.50 & 0.011088 & 0.049723 & 0.9215 & 0.78 & 0.8663 & 1.52 & 1.3048 & 2.59 & 0.7884 & 0.79 \\ 
      & 0.90 & 0.011015 & 0.049504 & 0.9153 & 0.70 & 0.8606 & 1.41 & 1.5682 & 5.63 & 0.7777 & 0.54 \\ 
 \hline 
30.00  & -0.90 & 0.006136 & 0.033517 & 0.9485 & 0.23 & 0.9088 & 0.53 & 0.5989 & 3.84 & 0.8601 & 0.47 \\ 
      & 0.00 & 0.006086 & 0.033333 & 0.9420 & 0.14 & 0.9026 & 0.55 & 0.9972 & 0.06 & 0.8476 & 0.06 \\ 
      & 0.90 & 0.006036 & 0.033152 & 0.9341 & 0.14 & 0.8952 & 0.42 & 1.4652 & 3.22 & 0.8323 & 0.002 \\ 
 \hline 
 \hline 
\end{tabular} 
\end{table*}


\twocolumngrid
  
\begin{table*}[!htbp]
\caption{Case $\ha =-0.9$. Compare caption of
  Table~\ref{tab:full_infos_a00} for general descriptions.
  Note the square brackets around all
  stated values concerning the multipolar fluxes for $\sigma\neq0$
  at $\hr=30$. These values in brackets refer to multipolar fluxes
  with respect to the $Y_{-2\lm}$ decomposition, whereas the
  table otherwise states values with respect
  to the $S_{-2\lm}$ basis. The reason is that at very large radii
  we did not compute the $S_{-2\lm}$-fluxes accurately,
  as discussed in Sec.~\ref{sub:nums}. The comparison with the analytical
  formulas, which refer to the $S_{-2\lm}$ decomposition,
  is nonetheless valid because for the stated modes,
  at such low frequencies, we verified that
  $ S_{-2\lm} \simeq Y_{-2\lm}$. }
\label{tab:full_infos_am9} 
\begin{tabular}[t]{| c | c c c | c c | c c | c c | c c |}
 \hline 
 \multicolumn{12}{|l|}{ {\bf\large{$\hat{a}=-0.90$}} } \\ \hline 
 $\hat{r}$ & $\sigma$ & $M\Omega$ & $x$ & 
 $\hat{F}_{m\leq3}$ & $\Delta \hat{F}_{m\leq3} [\%]$  & 
  $\hat{F}_{\rm{S22}}$ & $\Delta \hat{F}_{\rm{S22}} [\%] $ & 
 $\hat{F}_{\rm{S21}}$ & $\Delta \hat{F}_{\rm{S21}} [\%] $ & 
 $\hat{F}_{\rm{S33}}$ & $\Delta \hat{F}_{\rm{S33}} [\%] $ \\ 
 \hline 
5.00  & -0.90 & 0.117709 & 0.240182 & 6.4781 & 84.00 & 4.4357 & 94.93 & 13.2093 & 82.52 & 5.7381 & 64.47 \\ 
      & -0.50 & 0.107980 & 0.226757 & 4.6611 & 80.14 & 3.3073 & 92.63 & 10.7512 & 76.25 & 3.9652 & 55.31 \\ 
      & 0.00 & 0.097273 & 0.211509 & 3.1231 & 70.36 & 2.3101 & 88.77 & 8.4271 & 65.51 & 2.5068 & 40.50 \\ 
      & 0.50 & 0.088022 & 0.197877 & 2.1067 & 65.70 & 1.6213 & 83.45 & 6.6711 & 50.56 & 1.5675 & 19.55 \\ 
      & 0.90 & 0.081574 & 0.188091 & 1.5575 & 57.43 & 1.2332 & 77.96 & 5.6404 & 35.65 & 1.0789 & 2.54 \\ 
 \hline 
6.00  & -0.90 & 0.083178 & 0.190548 & 2.5692 & 57.56 & 1.9524 & 69.67 & 4.7113 & 60.49 & 2.2341 & 29.84 \\ 
      & -0.50 & 0.078170 & 0.182821 & 2.2155 & 54.79 & 1.7145 & 67.21 & 4.8787 & 54.96 & 1.8580 & 24.13 \\ 
      & 0.00 & 0.072480 & 0.173838 & 1.8252 & 45.13 & 1.4441 & 63.39 & 4.9193 & 46.51 & 1.4529 & 14.90 \\ 
      & 0.50 & 0.067380 & 0.165585 & 1.4840 & 44.41 & 1.1997 & 58.43 & 4.7967 & 35.89 & 1.1107 & 2.14 \\ 
      & 0.90 & 0.063698 & 0.159496 & 1.2557 & 38.76 & 1.0312 & 53.74 & 4.6663 & 26.39 & 0.8907 & 10.26 \\ 
 \hline 
7.00  & -0.90 & 0.063048 & 0.158409 & 1.7770 & 38.87 & 1.4272 & 48.27 & 3.0020 & 46.35 & 1.5407 & 14.43 \\ 
      & -0.50 & 0.060132 & 0.153487 & 1.6183 & 37.11 & 1.3152 & 46.66 & 3.3610 & 41.14 & 1.3550 & 10.54 \\ 
      & 0.00 & 0.056753 & 0.147681 & 1.4326 & 28.96 & 1.1805 & 44.10 & 3.7271 & 34.52 & 1.1454 & 4.61 \\ 
      & 0.50 & 0.053653 & 0.142253 & 1.2543 & 30.30 & 1.0474 & 40.56 & 3.9731 & 26.93 & 0.9528 & 3.49 \\ 
      & 0.90 & 0.051364 & 0.138178 & 1.1260 & 26.61 & 0.9494 & 37.27 & 4.1277 & 20.72 & 0.8195 & 10.92 \\ 
 \hline 
8.00  & -0.90 & 0.050040 & 0.135792 & 1.4629 & 26.85 & 1.2185 & 33.83 & 2.3078 & 37.00 & 1.2643 & 7.29 \\ 
      & -0.50 & 0.048194 & 0.132432 & 1.3683 & 25.72 & 1.1487 & 32.82 & 2.6880 & 31.89 & 1.1465 & 4.52 \\ 
      & 0.00 & 0.046025 & 0.128428 & 1.2557 & 19.06 & 1.0639 & 31.19 & 3.1322 & 26.58 & 1.0114 & 0.56 \\ 
      & 0.50 & 0.044003 & 0.124640 & 1.1424 & 21.22 & 0.9764 & 28.73 & 3.5031 & 21.06 & 0.8815 & 4.91 \\ 
      & 0.90 & 0.042488 & 0.121762 & 1.0584 & 18.76 & 0.9103 & 26.48 & 3.7752 & 16.95 & 0.7886 & 9.70 \\ 
 \hline 
10.00  & -0.90 & 0.034471 & 0.105918 & 1.2058 & 13.93 & 1.0499 & 17.92 & 1.7209 & 25.90 & 1.0416 & 2.12 \\ 
      & -0.50 & 0.033596 & 0.104117 & 1.1578 & 13.41 & 1.0122 & 17.49 & 2.0786 & 20.85 & 0.9771 & 0.55 \\ 
      & 0.00 & 0.032549 & 0.101944 & 1.1008 & 8.92 & 0.9667 & 16.84 & 2.5386 & 17.14 & 0.9023 & 1.43 \\ 
      & 0.50 & 0.031554 & 0.099855 & 1.0401 & 11.26 & 0.9174 & 15.55 & 2.9757 & 14.06 & 0.8261 & 4.37 \\ 
      & 0.90 & 0.030794 & 0.098245 & 0.9943 & 10.08 & 0.8798 & 14.45 & 3.3271 & 12.33 & 0.7703 & 6.74 \\ 
 \hline 
12.00  & -0.90 & 0.025653 & 0.086981 & 1.1021 & 7.94 & 0.9850 & 10.43 & 1.4703 & 19.75 & 0.9567 & 0.78 \\ 
      & -0.50 & 0.025171 & 0.085888 & 1.0719 & 7.67 & 0.9602 & 10.22 & 1.8006 & 14.82 & 0.9136 & 0.23 \\ 
      & 0.00 & 0.024589 & 0.084558 & 1.0364 & 4.50 & 0.9307 & 9.98 & 2.2403 & 11.97 & 0.8640 & 1.33 \\ 
      & 0.50 & 0.024028 & 0.083269 & 0.9969 & 6.52 & 0.8974 & 9.20 & 2.6788 & 10.17 & 0.8111 & 3.22 \\ 
      & 0.90 & 0.023595 & 0.082266 & 0.9673 & 5.88 & 0.8724 & 8.61 & 3.0449 & 9.67 & 0.7724 & 4.59 \\ 
 \hline 
15.00  & -0.90 & 0.018005 & 0.068697 & 1.0332 & 3.87 & 0.9458 & 5.26 & 1.2932 & 14.70 & 0.9066 & 0.37 \\ 
      & -0.50 & 0.017771 & 0.068098 & 1.0148 & 3.75 & 0.9303 & 5.20 & 1.5866 & 9.80 & 0.8786 & 0.26 \\ 
      & 0.00 & 0.017484 & 0.067364 & 0.9939 & 1.72 & 0.9121 & 5.19 & 1.9920 & 7.72 & 0.8471 & 0.77 \\ 
      & 0.50 & 0.017205 & 0.066647 & 0.9691 & 3.22 & 0.8903 & 4.73 & 2.4058 & 6.81 & 0.8116 & 1.94 \\ 
      & 0.90 & 0.016988 & 0.066084 & 0.9509 & 2.92 & 0.8743 & 4.45 & 2.7625 & 7.24 & 0.7855 & 2.71 \\ 
 \hline 
20.00  & -0.90 & 0.011504 & 0.050961 & 0.9905 & 1.42 & 0.9277 & 2.12 & 1.1811 & 11.57 & 0.8892 & 0.85 \\ 
      & -0.50 & 0.011410 & 0.050682 & 0.9802 & 1.37 & 0.9182 & 2.07 & 1.4296 & 6.55 & 0.8699 & 0.25 \\ 
      & 0.00 & 0.011294 & 0.050338 & 0.9694 & 0.27 & 0.9085 & 2.21 & 1.7768 & 4.40 & 0.8493 & 0.18 \\ 
      & 0.50 & 0.011180 & 0.049999 & 0.9545 & 1.18 & 0.8955 & 1.97 & 2.1721 & 5.20 & 0.8259 & 1.00 \\ 
      & 0.90 & 0.011090 & 0.049732 & 0.9444 & 1.07 & 0.8846 & 1.70 & 2.4992 & 6.13 & 0.8225 & 0.16 \\ 
 \hline
30.00  &  -0.90 & 0.006179 & 0.033672 & 0.9708 & 0.33 & [0.9291] & [0.59] & [1.067] & [5.92] & [0.8900] & [0.45] \\ 
       & 0.00 & 0.006119 & 0.033456 & 0.9607 & 0.23 & 0.9196 & 0.66 & 1.5795 & 2.00 & 0.8710 & 0.15 \\  
       &  0.90 & 0.006061 & 0.033243 & 0.9491 & 0.25  & [0.9086] & [0.55] & [2.148] & [2.80] & [0.8500] & [0.44] \\ 
 \hline 
 \hline 
\end{tabular} 
\end{table*}

\begin{table*}[!htbp]
\caption{Case $\ha =0.9$. Compare caption of
  Table~\ref{tab:full_infos_a00} for description.
  Additionally, in this table there appear 
  configurations that are, as yet, not covered by our experiments.
  These are marked with the $\times$-symbol.
  Furthermore, note the square brackets around all
  stated values concerning the multipolar fluxes for $\sigma\neq0$
  at $\hr=30$. These values in brackets refer to multipolar fluxes
  with respect to the $Y_{-2\lm}$ decomposition, whereas the
  table otherwise states values with respect
  to the $S_{-2\lm}$ basis. The reason is that at very large radii
  we did not compute the $S_{-2\lm}$-fluxes accurately,
  as discussed in Sec.~\ref{sub:nums}. The comparison with the analytical
  formulas, which refer to the $S_{-2\lm}$ decomposition,
  is nonetheless valid because for the stated modes,
  at such low frequencies, we verified that
  $ S_{-2\lm} \simeq Y_{-2\lm}$. }
\label{tab:full_infos_ap9} 
 \begin{tabular}[t]{| c | c c c | c c | c c | c c | c c |}
 \hline 
 \multicolumn{12}{|l|}{ {\bf\large{$\hat{a}=0.90$}} } \\ \hline 
 $\hat{r}$ & $\sigma$ & $\Omega$ & $x$ & 
 $\hat{F}_{m\leq3}$ & $\Delta \hat{F}_{m\leq3} [\%]$  & 
  $\hat{F}_{\rm{S22}}$ & $\Delta \hat{F}_{\rm{S22}} [\%] $ & 
 $\hat{F}_{\rm{S21}}$ & $\Delta \hat{F}_{\rm{S21}} [\%] $ & 
 $\hat{F}_{\rm{S33}}$ & $\Delta \hat{F}_{\rm{S33}} [\%] $ \\ 
 \hline 
4.00  & -0.90 & 0.124009 & 0.248677 & 0.7725 & 124.37 & 0.6023 & 224.00 & 0.0004 & 469114.18 & 0.5038 & 145.45 \\ 
      & -0.50 & 0.118248 & 0.240914 & 0.7657 & 111.89 & 0.6063 & 193.82 & 0.0311 & 3703.59 & 0.4873 & 116.16 \\ 
      & 0.00 & 0.112360 & 0.232848 & 0.7561 & 112.04 & 0.6070 & 164.26 & 0.1451 & 334.74 & 0.4667 & 82.26 \\ 
      & 0.50 & 0.107776 & 0.226472 & 0.7427 & 89.70 & 0.6024 & 141.41 & 0.3058 & 47.12 & 0.4443 & 50.82 \\ 
      & 0.90 & 0.104933 & 0.222472 & 0.7355 & 83.57 & 0.6001 & 126.34 & 0.4618 & 137.58 & 0.4299 & 24.94 \\ 
 \hline 
6.00  & -0.90 & 0.068067 & 0.166708 & 0.8200 & 38.65 & 0.6886 & 65.27 & $\times$ & $\times$ & 0.5809 & 52.87 \\ 
      & -0.50 & 0.066199 & 0.163645 & 0.8125 & 35.78 & 0.6866 & 58.84 & $\times$ & $\times$ & 0.5654 & 41.44 \\ 
      & 0.00 & 0.064115 & 0.160192 & 0.8047 & 35.16 & 0.6841 & 51.74 & 0.2229 & 110.39 & 0.5478 & 27.08 \\ 
      & 0.50 & 0.062293 & 0.157142 & 0.7933 & 29.68 & 0.6777 & 45.23 & $\times$ & $\times$ & 0.5277 & 13.59 \\ 
      & 0.90 & 0.061011 & 0.154978 & 0.7860 & 27.66 & 0.6736 & 40.61 & 0.6288 & 61.86 & 0.5137 & 2.36 \\ 
 \hline 
8.00  & -0.90 & 0.044299 & 0.125197 & 0.8447 & 16.51 & 0.7370 & 27.42 & 0.0320 & 1329.22 & 0.6339 & 25.20 \\ 
      & -0.50 & 0.043468 & 0.123627 & 0.8385 & 15.42 & 0.7339 & 25.08 &  $\times$ & $\times$ & 0.6205 & 19.11 \\ 
      & 0.00 & 0.042504 & 0.121792 & 0.8325 & 14.81 & 0.7310 & 22.47 & 0.2835 & 52.23 & 0.6059 & 11.21 \\ 
      & 0.50 & 0.041619 & 0.120096 & 0.8229 & 12.94 & 0.7246 & 19.71 &  $\times$ & $\times$  & 0.5881 & 3.97 \\ 
      & 0.90 & 0.040966 & 0.118837 & 0.8169 & 12.05 & 0.7205 & 17.74 & 0.7217 & 38.38 & 0.5758 & 2.15 \\ 
 \hline 
10.00   & -0.90 & 0.031712 & 0.100188 & 0.8618 & 8.36 & 0.7701 & 13.96 & 0.0590 & 433.26 & 0.6736 & 13.97 \\ 
        & -0.50 & 0.031271 & 0.099256 & 0.8566 & 7.83 & 0.7670 & 12.85 &  $\times$ & $\times$ & 0.6621 & 10.22 \\ 
        & 0.00 & 0.030748 & 0.098147 & 0.8520 & 7.33 & 0.7643 & 11.69 & 0.3319 & 29.76 & 0.6499 & 5.20 \\ 
        & 0.50 & 0.030256 & 0.097097 & 0.8438 & 6.57 & 0.7582 & 10.20 &  $\times$ & $\times$  & 0.6342 & 0.80 \\ 
        & 0.90 & 0.029884 & 0.096300 & 0.8388 & 6.10 & 0.7545 & 9.19 & 0.7794 & 27.41 & 0.6234 & 3.00 \\ 
 \hline 
12.00   & -0.90 & 0.024124 & 0.083489 & 0.8749 & 4.70 & 0.7948 & 8.00 &   $\times$ & $\times$  & 0.7048 & 8.55 \\ 
        & -0.50 & 0.023861 & 0.082883 & 0.8705 & 4.40 & 0.7919 & 7.40 &  $\times$ & $\times$  & 0.6948 & 6.02 \\ 
        & 0.00 & 0.023546 & 0.082152 & 0.8669 & 4.00 & 0.7895 & 6.83 & 0.3716 & 18.98 & 0.6846 & 2.53 \\ 
        & 0.50 & 0.023246 & 0.081451 & 0.8597 & 3.67 & 0.7839 & 5.89 &   $\times$ & $\times$ & 0.6705 & 0.37 \\ 
        & 0.90 & 0.023015 & 0.080911 & 0.8554 & 3.39 & 0.7805 & 5.30 & 0.8184 & 21.18 & 0.6613 & 2.99 \\ 
 \hline 
15.00   & -0.90 & 0.017257 & 0.066780 & 0.8900 & 2.23 & 0.8225 & 4.01 & 0.1257 & 70.13 & 0.7412 & 4.63 \\ 
        & -0.50 & 0.017119 & 0.066423 & 0.8864 & 2.08 & 0.8199 & 3.71 &  $\times$ & $\times$  & 0.7326 & 3.11 \\ 
        & 0.00 & 0.016951 & 0.065987 & 0.8839 & 1.79 & 0.8181 & 3.53 & 0.4197 & 11.06 & 0.7248 & 0.80 \\ 
        & 0.50 & 0.016788 & 0.065564 & 0.8777 & 1.70 & 0.8130 & 2.96 &  $\times$ & $\times$  & 0.7127 & 0.89 \\ 
        & 0.90 & 0.016662 & 0.065235 & 0.8743 & 1.55 & 0.8100 & 2.64 & 0.8574 & 15.69 & 0.7046 & 2.46 \\ 
 \hline 
20.00   & -0.90 & 0.011204 & 0.050069 & 0.9077 & 0.75 & 0.8552 & 1.71 & 0.1830 & 17.94 & 0.7862 & 1.83 \\ 
        & -0.50 & 0.011143 & 0.049889 & 0.9051 & 0.68 & 0.8519 & 1.45 & 0.3051 & 13.03 & 0.7799 & 0.98 \\ 
        & 0.00 & 0.011069 & 0.049667 & 0.9037 & 0.53 & 0.8509 & 1.50 & 0.4797 & 5.60 & 0.7723 & 0.11 \\ 
        & 0.50 & 0.010996 & 0.049450 & 0.8986 & 0.51 & 0.8461 & 1.11 & 0.7000 & 4.49 & 0.7631 & 0.98 \\ 
        & 0.90 & 0.010939 & 0.049279 & 0.8961 & 0.45 & 0.8444 & 1.05 & 0.9177 & 13.00 & 0.7577 & 1.93 \\ 
 \hline 
30.00  & -0.90 & 0.006094 & 0.033365 & 0.9306 & 0.16 & [0.8924] & [0.46] & [0.2722] & [0.64] & [0.8381] & [0.52] \\ 
       & 0.00 & 0.006053 & 0.033212 & 0.9275 & 0.02 & 0.8896 & 0.45 & 0.5585 & 2.20 & 0.8296 & 0.36 \\ 
       & 0.90 & 0.006012 & 0.033062 & 0.9234 & 0.08  & [0.8856] & [0.31] & [0.9262] & [5.95] & [0.8190] & [0.99] \\ 
 \hline 
 \hline 
\end{tabular} 
\end{table*}


\bibliographystyle{unsrt}
\bibliography{../../refs/refs}

\end{document}